\documentclass[letterpaper,journal,twoside,final,onecolumn,12pt]{IEEEtran}

\usepackage{bm}
\usepackage{cite}

\newtheorem{thm}{\protect\theoremname}
\newtheorem{lem}{\protect\lemmaname}
\providecommand{\lemmaname}{Lemma}
\providecommand{\theoremname}{Theorem}

\ifCLASSINFOpdf
  \usepackage[pdftex]{graphicx}
\else
  \usepackage[dvips]{graphicx}
\fi

\usepackage[cmex10]{amsmath}
\usepackage{amssymb}

\usepackage{algorithm}
\usepackage{algpseudocode}

\usepackage{fixltx2e}
\begin{document}
%
\title{Compute-Compress-and-Forward: Exploiting Asymmetry of Wireless Relay Networks}

\author{Yihua~Tan~and~Xiaojun~Yuan,~\IEEEmembership{Member,~IEEE}
\thanks{Yihua Tan is with Institute of Network Coding, The Chinese University of Hong Kong, Hong Kong, China. 
E-mail: ty013@ie.cuhk.edu.hk}
\thanks{Xiaojun Yuan is with School of Information Science and Technology, ShanghaiTech University, Shanghai, China.
E-mail: yuanxj@shanghaitech.edu.cn}
}

\markboth{IEEE Transactions on Signal Processing}{Tan Compute-Compress-and-Forward}

\maketitle

\begin{abstract}
Compute-and-forward (CF) harnesses interference in a wireless network
by allowing relays to compute combinations of source messages. The
computed message combinations at relays are correlated, and so directly
forwarding these combinations to a destination generally incurs information
redundancy and spectrum inefficiency. To address this issue, we propose
a novel relay strategy, termed \textit{compute-compress-and-forward}
(CCF). In CCF, source messages are encoded using nested lattice codes
constructed on a chain of nested coding and shaping lattices. A key
difference of CCF from CF is an extra compressing stage inserted in
between the computing and forwarding stages of a relay, so as to reduce
the forwarding information rate of the relay. The compressing stage
at each relay consists of two operations: first to quantize the computed
message combination on an appropriately chosen lattice (referred to
as a quantization lattice), and then to take modulo on another lattice
(referred to as a modulo lattice). We study the design of the quantization
and modulo lattices and propose successive recovering algorithms to
ensure the recoverability of source messages at destination. Based
on that, we formulate a sum-rate maximization problem that is in general
an NP-hard mixed integer program. A low-complexity algorithm is proposed
to give a suboptimal solution. Numerical results are presented to
demonstrate the superiority of CCF over the existing CF schemes.
\end{abstract}


%
\IEEEpeerreviewmaketitle

\section{Introduction}

\IEEEPARstart{I}{nterference} has been long regarded as an adverse
factor for wireless communications until the seminal
work of compute-and-forward (CF) \cite{nazer2011compute}. The main
idea of CF is to harness interference by allowing relays to compute
linear combinations of source messages, without even the knowledge
of any individual source messages. 
Since the advent of CF in \cite{nazer2011compute},
various CF-based schemes have been investigated in the literature. 
Much progress has been made towards the understanding of fundamental
characterizations of wireless relay networks 
\cite{niesen2012degrees,osmane2011compute,ntranos2013usc,zhu2014asymmetric,fang2014asymptotic,yuanmultiwayrelay,yuan2013mimo,wang2013weighted,wangmultiwaydof,yihua2014asymmetric,hern2013multilevel,huang2013opportunistic,huang2013compute,yang2012achievable,gunduz2013multiway,lee2012achievable,tian2013degrees}. 

A CF scheme usually employs nested lattice codes \cite{nam2011nested,ordentlich2012simple,erez2004achieving,bannai1999sphere},
where the codebook of a nested lattice code is defined as the lattice
points of a coding lattice confined within the fundamental Voronoi
region of a nested shaping lattice \cite{erez2004achieving,feng2013algebraic}.
In the original CF scheme \cite{nazer2011compute}, a common shaping
lattice is assumed for all source nodes, implying that every source
node is forced to use a common power for transmission. This limits
the potential of a CF scheme to exploit the asymmetry inherent in
the nature of wireless communication channels. 

Recent work in \cite{ntranos2013usc,zhu2014asymmetric} presented
modified CF schemes with asymmetrically constructed lattice codes,
in which not only coding lattices but also shaping lattices are chosen
from a chain of nested lattices. Precoding techniques based on channel
state information (CSI) were proposed to enhance the performance of
asymmetric CF schemes \cite{zhu2014asymmetric}. However, both \cite{ntranos2013usc}
and \cite{zhu2014asymmetric} were focused on asymmetry in the first
hop of wireless relay networks. Not well understood is how the relays
should process their received message combinations, or more specifically,
how they should optimize the overall system when multiple hops of
the network exhibit asymmetry.

In a multi-hop relay network, the message combinations computed at
relays are generally correlated, as they are generated from a common
set of source messages. This implies that directly forwarding these
combinations in general leads to information redundancy at destination.
Meanwhile, the forwarding channel seen by each relay may vary significantly
from each other due to the asymmetry of channel fading. 
It is thus desirable to reduce the forwarding rates of the relays with 
relatively bad channel quality. 
This inspires us to seek for more efficient relaying techniques
for multi-hop relay networks.

In this paper, we propose a novel relay strategy, termed compute-compress-and-forward
(CCF). A key difference of CCF from CF, as manifested by their names,
is an extra compressing stage inserted in between the computing and
forwarding stages of each relay. The compressing stage at a relay
consists of two operations: first to quantize the computed message
combination on a lattice (referred to as a \textit{quantization lattice}),
and then to take modulo on another lattice (referred to as a \textit{modulo
lattice}). 
The design of the quantization and modulo operation should
take into account the following two aspects. 
On one hand, it is desirable to choose a quantization lattice 
as coarse as possible and a modulo lattice as fine as possible, 
so as to minimize the forwarding rate at each relay. 
On the other hand, quantization and modulo operation in
general suffer from information loss, and so the design of these compressing
operations should be subject to the recoverability of source messages
at destination. As such, there is a balance to strike in the design
of the quantization and modulo lattices for compressing.

To concretize the idea of CCF, we consider a two-hop relay network
with multiple sources, multiple relays, and a single destination.
The quantization and modulo lattices for compressing are respectively
chosen as permutations of the coding and shaping lattices used for
source coding. 
We present successive recovering algorithms to recover source messages at the destination. 
We then show that the above choice of compressing lattices is optimal in the sense of
minimizing the forwarding sum rate under the constraint of the recoverability
of source messages at the destination. Based on that, we formulate
a sum-rate maximization problem that is generally an NP-hard mixed
integer program. We propose a low-complexity suboptimal solution to
this problem by utilizing the Lenstra\textendash Lenstra\textendash Lovász
(LLL) lattice basis reduction algorithm \cite{osmane2011compute,sakzad2012integer}.
Numerical results are presented to demonstrate the superiority of
CCF over CF by exploiting the channel asymmetry of wireless networks.

This paper is organized as follows. In Section \ref{sec:Preliminaries},
we introduce the system model and some fundamentals of lattice and
nested lattice codes. In Section \ref{sec:ACF}, we describe asymmetric
CF for the first hop of the considered network. In Section \ref{sec:Forwarding},
we describe the proposed CCF scheme involving quantization and modulo
operation at relays. Section \ref{sec:Modulo-Operation-at-the-relays}
is focused on the design of modulo operation, and Section \ref{sec:Quantization-at-the-relays}
on quantization. The joint design of quantization and modulo operation
is investigated in Section \ref{sec:Asymmetric-Quan-Mod}. In Section
\ref{sec:Performance-Optimization}, we study the sum-rate maximization
problem for the overall CCF scheme, and present numerical results
to demonstrate the advantage of CCF over CF. Finally, the concluding
remarks are presented in Section \ref{sec:Conclusions}.

\section{Preliminaries\label{sec:Preliminaries}}

\subsection{System Model}

\begin{figure}[tbh]
\centering
\includegraphics[width=0.75\columnwidth]{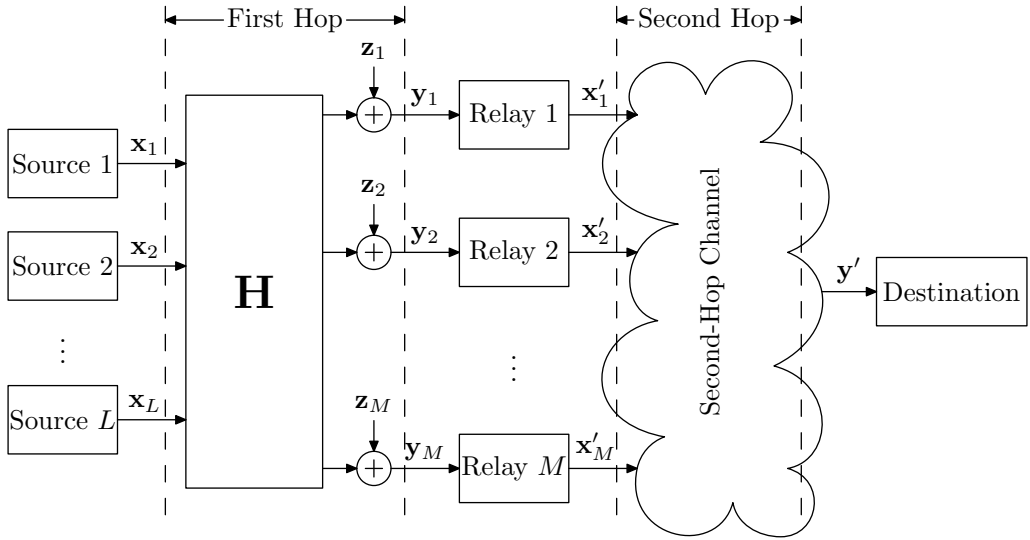}
\caption{A diagram representation of the system model.}
\label{fig:system-model}
\end{figure}

Consider a relay network in which $L$ source nodes transmit private
messages to a common destination via $M$ intermediate relay nodes.
Each source node is equipped with a single antenna, and so is each
relay node. Assume that there is no direct link between any source
node and the destination. A two-hop relay protocol is employed. In
the first hop, the source nodes transmit signals simultaneously to
the relay nodes. In the second hop, the relay nodes transmit signals
to the destination. The overall system model is illustrated in Fig.
\ref{fig:system-model}. 

In the first hop, each source has a message $\mathbf{w}_{l}\in\mathbb{F}_{\gamma}^{k_{l}}$,
where $\mathbb{F}_{\gamma}$ is a finite field of size $\gamma$ and
$\gamma$ is a prime number. Each source encodes $\mathbf{w}_{l}$
as $\mathbf{x}_{l}=f\left(\mathbf{w}_{l}\right)\in\mathbb{R}^{n_{1}\times1}$
and then transmit $\mathbf{x}_{l}$ in the first-hop channel. The
first-hop channel is a real Gaussian channel with additive white Gaussian
noise (AWGN), represented as 
\begin{equation}
\mathbf{y}_{m}=\sum_{l=1}^{L}h_{ml}\mathbf{x}_{l}+\mathbf{z}_{m},m=1,\cdots,M\label{eq:first-hop-model}
\end{equation}
where $\mathbf{y}_{m}\in\mathbb{R}^{n_{1}\times1}$ is the received
signal of the $m$-th relay, $h_{ml}\sim\mathcal{N}\left(0,1\right)$
is the channel coefficient of the link from source $l$ to relay $m$,
and $\mathbf{z}_{m}\in\mathbb{R}^{n_{1}\times1}$ is a Gaussian noise
vector drawn from $\mathcal{N}\left(0,\mathbf{I}_{n_{1}}\right)$.
Denote by $p_{l}=\frac{1}{n_{1}}\left\Vert \mathbf{x}_{l}\right\Vert ^{2}$
the average power of source $l$. Then, the power constraint of source
$l$ is given by 
\begin{equation}
p_{l}\leq P_{l}\label{eq:power_constraint}
\end{equation}
where $P_{l}$ is the power budget of source $l$. Further denote
by $\mathbf{H}=\left[h_{ml}\right]$ the first-hop channel matrix
and by $\mathbf{h}_{m}=\left[h_{m1},\cdots,h_{mL}\right]^{T}$ the
channel vector to the $m$-th relay. 

In the second hop, each relay $m$ communicates $\mathbf{x}_{m}^{\prime}\in\mathbb{R}^{n_{2}\times1}$
to the destination. The second-hop channel is defined by the transfer
probability density function $p\left(\mathbf{y}^{\prime}\vert\mathbf{x}_{1}^{\prime},\cdots,\mathbf{x}_{M}^{\prime}\right)$,
where $\mathbf{y}^{\prime}$ is the received signal at the destination.
The destination computes $\left\{ \hat{\mathbf{w}}_{l}\right\} $ as
an estimate of the original messages $\left\{ \mathbf{w}_{l}\right\} $.
Note that a detailed model of the second-hop channel is irrelevant
to most discussions in this paper. Thus, we will only give
an example of the second-hop channel later in Section \ref{sec:Performance-Optimization}. 

For convenience of discussion, we henceforth assume $n_{1}=n_{2}=n$
and $L=M$, i.e., the two hops have equal time duration and the number
of the source nodes is equal to the number of relay nodes.

We say that a rate tuple $\left(r_{1},r_{2},\cdots,r_{L}\right)$
is achievable if 
\[
\Pr\left(\hat{\mathbf{w}}_{l}=\mathbf{w}_{l},l=1,\cdots,L\right)\to0\;\textrm{as}\; n\to\infty
\]
i.e., the destination can reliably recover the original messages $\left\{ \mathbf{w}_{l}\right\} $
by $\mathbf{y}^{\prime}$ with a vanishing error probability as $n\to\infty$.
This paper aims to analyze the performance of the network described
above with CF-based relaying.

\subsection{Lattice and Nested Lattice Codes}

Nested lattice coding is a key technique used in CF-based relaying.
To set the stage for further discussion, we introduce some basic properties
of nested lattice codes. A lattice $\Lambda\subset\mathbb{R}^{n}$
is a discrete group under the addition operation, and can be represented
as 
\[
\Lambda=\left\{ \mathbf{s}=\mathbf{G}\mathbf{c}:\mathbf{c}\in\mathbb{Z}^{n}\right\} 
\]
where $\mathbf{G}\in\mathbb{R}^{n\times n}$ is a lattice generating
matrix \cite{ordentlich2012simple}. Let $\mathcal{V}$ denote the
fundamental Voronoi region of $\Lambda$. Every $\mathbf{x}\in\mathbb{R}^{n}$
can be uniquely written as $\mathbf{x}=Q_{\Lambda}\left(\mathbf{x}\right)+\mathbf{r}$,
where $\mathbf{r}\in\mathcal{V}$ and $Q_{\Lambda}\left(\mathbf{x}\right)$
is the quantization of $\mathbf{x}$ on $\Lambda$, i.e., the nearest
lattice point of $\mathbf{x}$ in $\Lambda$. Modulo-$\Lambda$ operation
\cite{erez2004achieving} is defined as 
\begin{equation}
\mathbf{x}\bmod\Lambda=\mathbf{x}-Q_{\Lambda}\left(\mathbf{x}\right).\label{eq:mod-quantization-relation}
\end{equation}
The second moment per dimension is defined as $\sigma^{2}\left(\mathcal{V}\right)\triangleq\frac{1}{n}\frac{\int_{\mathcal{V}}\left\Vert \mathbf{x}\right\Vert ^{2}dx}{\textrm{Vol}\left(\mathcal{V}\right)}$,
where $\textrm{Vol}\left(\mathcal{V}\right)$ is the volume of $\mathcal{V}$.
The normalized second moment of $\Lambda$ is defined as $G\left(\Lambda\right)\triangleq\frac{\sigma^{2}\left(\mathcal{V}\right)}{\left(\textrm{Vol}\left(\mathcal{V}\right)\right)^{2/n}}$.
We say that $\Lambda$ is good for MSE quantization \cite{erez2004achieving}
if 
\begin{equation}
\lim_{n\to\infty}G\left(\Lambda\right)=\frac{1}{2\pi e}\label{eq:good-for-MSE}
\end{equation}
where $e$ is the Euler's number. 

A lattice $\Lambda_{1}$ is nested in a lattice $\Lambda_{2}$ if
$\Lambda_{1}\subseteq\Lambda_{2}$. In this case, we say that $\Lambda_{1}$
is coarser than $\Lambda_{2}$, or $\Lambda_{2}$ is finer than $\Lambda_{1}$.
Further, if $\Lambda_{1}\subseteq\Lambda_{2}$, then for $\mathbf{x}\in\mathbb{R}^{n}$,
\begin{equation}
\left[\mathbf{x}\bmod\Lambda_{1}\right]\bmod\Lambda_{2}=\mathbf{x}\bmod\Lambda_{2}.\label{eq:nested-lattice-mod-relation}
\end{equation}

A lattice codebook can be represented using a nested lattice pair
$\left(\Lambda_{c},\Lambda_{s}\right)$ with $\Lambda_{s}\subseteq\Lambda_{c}$,
where $\Lambda_{s}$ is referred to as a shaping lattice and $\Lambda_{c}$
as a coding lattice. Denote the Voronoi regions of $\Lambda_{c}$
and $\Lambda_{s}$ respectively by $\mathcal{V}_{c}$ and $\mathcal{V}_{s}$,
and the corresponding volumes by $V_{c}$ and $V_{s}$. The generated
lattice codebook is 
\begin{equation}
\mathcal{C}=\Lambda_{c}\bmod\Lambda_{s}\triangleq\Lambda_{c}\cap\mathcal{V}_{s}.
\end{equation}
The rate of this nested lattice code is given by 
\begin{equation}
R=\frac{1}{n}\log\left|\mathcal{C}\right|=\frac{1}{n}\log\frac{V_{s}}{V_{c}}.\label{eq:rate_nested_lattice_code}
\end{equation}
Moreover, we say that $\Lambda_{1},\Lambda_{2},\cdots,\Lambda_{K}$
form a nested lattice chain if $\Lambda_{1}\supseteq\Lambda_{2}\supseteq\cdots\supseteq\Lambda_{K}$
\cite{nam2010capacity}. Nested lattice codes with various rates can
be constructed by appropriately selecting a pair of shaping and coding
lattices from the chain, as detailed in Section \ref{sec:ACF}.

\section{First Hop: Asymmetric Compute-and-Forward\label{sec:ACF}}

\begin{figure*}[tbh]
\centering
\includegraphics[width=1\textwidth]{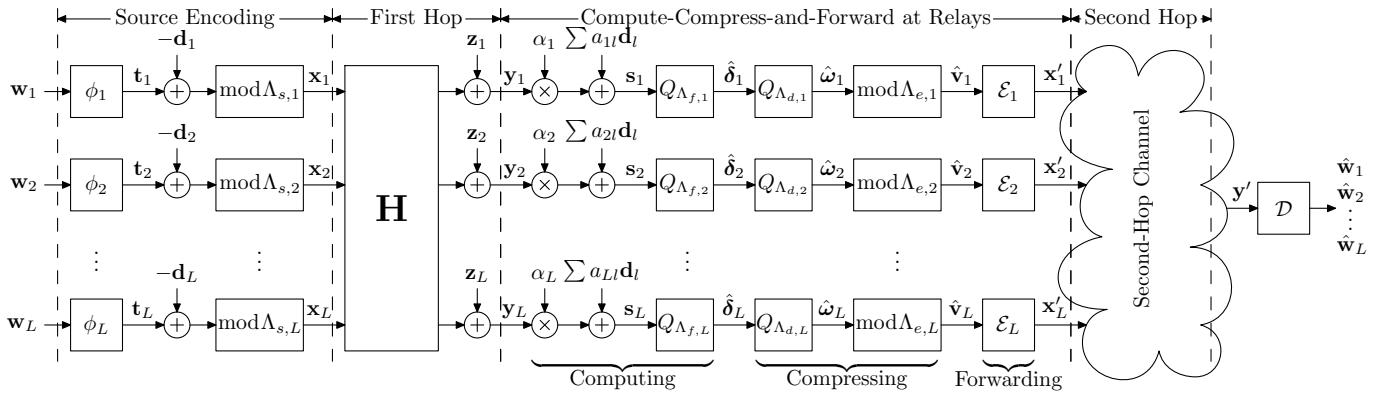}
\caption{The transceiver and relay operations
for the two-hop relay network in Fig. \ref{fig:system-model}.}
\label{fig:Compute-and-forward.}
\end{figure*}

In this paper, we propose CCF for a two-hop relay channel, as illustrated
in Fig. \ref{fig:Compute-and-forward.}. This section is focused on
the first hop which basically follows the asymmetric compute-and-forward
(ACF) in \cite{ntranos2013usc,zhu2014asymmetric}. The ACF scheme
involves asymmetric lattice coding with nested coding and shaping
lattices, which improves performance by exploiting the knowledge of
CSI.

\subsection{\label{sub:encoding-source}Encoding at Sources}

We use nested lattice codes to encode the messages of the sources.
The lattices are generated following Construction A method in \cite{nazer2011compute,bannai1999sphere}.
Let $\kappa\left(\cdot\right)$ be the mapping from the prime-sized
finite field $\mathbb{F}_{\gamma}$ to the corresponding integers
$\left\{ 0,1,\cdots,\gamma-1\right\} $, and $\kappa^{-1}\left(\cdot\right)$
be the inverse mapping of $\kappa\left(\cdot\right)$. Note that $\kappa\left(\cdot\right)$
can be applied to a vector or matrix in an entry-wise manner.

We first construct a chain of nested coding lattices following the
Construction A method in \cite{nazer2011compute,bannai1999sphere}.
Let $\mathbf{G}\in\mathbb{F}_{\gamma}^{n\times k}$ be a random matrix
with i.i.d. elements uniformly drawn over $\mathbb{F}_{\gamma}$.
Denote by $\mathbf{G}_{A,l}$ the first $k_{A,l}$ columns of $\mathbf{G}$,
where $k_{A,l}\leq k$ is an integer. Define $\mathcal{L}_{A,l}=\left\{ \mathbf{G}_{A,l}\mathbf{b}:\mathbf{b}\in\mathbb{F}_{\gamma}^{k_{A,l}}\right\} $,
and construct a lattice $\tilde{\Lambda}_{A,l}=\gamma^{-1}\kappa\left(\mathcal{L}_{A,l}\right)+\mathbb{Z}^{n}$.
Finally, construct a coding lattice as $\Lambda_{A,l}=\mathbf{B}\tilde{\Lambda}_{A,l}$,
where $\mathbf{B}\in\mathbb{R}^{n\times n}$ is a lattice generation
matrix. In the construction, we require $k_{A,1}\geq k_{A,2}\ge\cdots\geq k_{A,L}$,
and so the constructed lattices are nested as $\Lambda_{A,1}\supseteq\Lambda_{A,2}\supseteq\cdots\supseteq\Lambda_{A,L}$.
Similarly, we construct a chain of nested shaping lattices $\Lambda_{B,1}\supseteq\Lambda_{B,2}\supseteq\cdots\supseteq\Lambda_{B,L}$,
based on the same matrices $\mathbf{G}$ and $\mathbf{B}$, with parameters
$k_{B,1}\geq k_{B,2}\ge\cdots\geq k_{B,L}$.

The lattice codebook $\mathcal{C}_{l}$ for each source $l$ is constructed
as follows. We designate for the $l$-th source a coding lattice $\Lambda_{c,l}=\Lambda_{A,\pi_{c}\left(l\right)}$,
and a shaping lattice $\Lambda_{s,l}=\Lambda_{B,\pi_{s}\left(l\right)}$,
where permutations $\pi_{c}\left(\cdot\right)$ and $\pi_{s}\left(\cdot\right)$
are bijective mappings from $\left\{ 1,\cdots,L\right\} $ to $\left\{ 1,\cdots,L\right\} $.
In constructing lattice codebooks, we require that $\Lambda_{s,l}$
is nested in $\Lambda_{c,l}$, and thus, $k_{B,\pi_{s}\left(l\right)}<k_{A,\pi_{c}\left(l\right)}$.
Then the codebook of source $l$ is $\mathcal{C}_{l}=\Lambda_{c,l}\cap\mathcal{V}_{s,l}$.
Note that the coding lattices $\left\{ \Lambda_{c,l}\right\} $ and
shaping lattices $\left\{ \Lambda_{s,l}\right\} $ constructed above
are good for both AWGN \cite{nazer2011compute} and mean square error
(MSE) quantization \cite{erez2004achieving}.

We are now ready to describe the encoding function at each source.
Let $k_{c,l}=k_{A,\pi_{c}\left(l\right)}$, and $k_{s,l}=k_{B,\pi_{s}\left(l\right)}$.
The $l$-th source draws a vector $\tilde{\mathbf{w}}_{l}$ over $\mathbb{F}_{\gamma}$
with length $\left(k_{c,l}-k_{s,l}\right)$, and zero-pad $\tilde{\mathbf{w}}_{l}$
to form a message as 
\begin{equation}
\mathbf{w}_{l}=\bigl[\underbrace{0,\cdots,0}_{k_{s,l}},\underbrace{\tilde{\mathbf{w}}_{l}^{T}}_{k_{c,l}-k_{s,l}},\underbrace{0,\cdots,0}_{k-k_{c,l}}\bigl]^{T}\in\mathbb{F}_{\gamma}^{k\times1}.
\end{equation}
The $l$-th source maps $\mathbf{w}_{l}$ to a lattice codeword in
$\mathcal{C}_{l}$ as 
\begin{equation}
\mathbf{t}_{l}=\phi_{l}\left(\mathbf{w}_{l}\right)\triangleq\left[\mathbf{B}\gamma^{-1}\kappa\left(\mathbf{G}\mathbf{w}_{l}\right)\right]\bmod\Lambda_{s,l}.\label{eq:message-lattice-mapping}
\end{equation}
By following the proof of Lemma 5 in \cite{nazer2011compute}, it
can be shown that $\phi_{l}\left(\cdot\right)$ is a one-to-one mapping,
which gives an isomorphism between the finite-field codebook $\mathbb{F}_{\gamma}^{k_{c,l}-k_{s,l}}$
and lattice codebook $\mathcal{C}_{l}$. Then, we construct the signal
as 
\begin{equation}
\mathbf{x}_{l}=\left(\mathbf{t}_{l}-\mathbf{d}_{l}\right)\bmod\Lambda_{s,l},\label{eq:x-transmit-signal}
\end{equation}
where $\mathbf{d}_{l}\in\mathbb{R}^{n\times1}$ is a random dithering
signal uniformly distributed in the Voronoi region $\mathcal{V}_{s,l}$
of $\Lambda_{s,l}$. From Lemma 1 of \cite{erez2004achieving}, $\mathbf{x}_{l}$
is uniformly distributed over $\mathcal{V}_{s,l}$. Then, the average
power of $\mathbf{x}_{l}$ is given by 
\begin{equation}
p_{l}=\textrm{E}\left[\frac{1}{n}\left\Vert \mathbf{x}_{l}\right\Vert ^{2}\right]=G\left(\Lambda_{s,l}\right)\left(\textrm{Vol}\left(\mathcal{V}_{s,l}\right)\right)^{\frac{2}{n}}\leq P_{l}\label{eq:lattice_average_power}
\end{equation}
where $\textrm{E}\left(x\right)$ denotes the expectation of $x$,
and $P_{l}$ is the power budget of source $l$ in \eqref{eq:power_constraint}.

\subsection{\label{sub:Computing-relays}Computing at Relays}

We now consider the relay operations. From \eqref{eq:first-hop-model}
and \eqref{eq:x-transmit-signal}, each relay $m$ receives 
\begin{equation}
\mathbf{y}_{m}=\sum_{l=1}^{L}h_{ml}\left(\mathbf{t}_{l}-\mathbf{d}_{l}\right)\bmod\Lambda_{s,l}+\mathbf{z}_{m}
\label{eq:y_m}
\end{equation}
and computes a linear combination: 
\begin{equation}
\bm{\delta}_{m}\triangleq\sum_{l=1}^{L}a_{ml}\left(\mathbf{t}_{l}-Q_{\Lambda_{s,l}}\left(\mathbf{t}_{l}-\mathbf{d}_{l}\right)\right)\label{eq:omega-def}
\end{equation}
where $a_{ml},l=1,\cdots,L$, are integer coefficients. To this end,
the relay first multiplies $\mathbf{y}_{m}$ by $\alpha_{m}$ and
removes the dithering signals, yielding
\begin{eqnarray}
\mathbf{s}_{m} & = & \alpha_{m}\mathbf{y}_{m}+\sum_{l=1}^{L}a_{ml}\mathbf{d}_{l}\nonumber \\
 & \overset{\left(a\right)}{=} & \sum_{l=1}^{L}\left(a_{ml}\left(\mathbf{x}_{l}+\mathbf{d}_{l}\right)+\theta_{ml}\mathbf{x}_{l}\right)+\alpha_{m}\mathbf{z}_{m}\nonumber \\
 & \overset{\left(b\right)}{=} & \sum_{l=1}^{L}\left(a_{ml}\left(\mathbf{t}_{l}-Q_{\Lambda_{s,l}}\left(\mathbf{t}_{l}-\mathbf{d}_{l}\right)\right)+\theta_{ml}\mathbf{x}_{l}\right)+\alpha_{m}\mathbf{z}_{m}\nonumber \\
 & = & \bm{\delta}_{m}+\sum_{l=1}^{L}\theta_{ml}\mathbf{x}_{l}+\alpha_{m}\mathbf{z}_{m}\label{eq:s_m-expression}
\end{eqnarray}
where step $\left(a\right)$ follows by \eqref{eq:y_m} and the definition of 
$\theta_{ml}\triangleq\alpha_{m}h_{ml}-a_{ml}$; 
step $\left(b\right)$ follows from \eqref{eq:mod-quantization-relation}
and \eqref{eq:x-transmit-signal}. Then the relay decodes $\bm{\delta}_{m}$
by quantizing $\mathbf{s}_{m}$ over a quantization lattice $\Lambda_{f,m}$,
yielding 
\begin{equation}
\hat{\bm{\delta}}_{m}=Q_{\Lambda_{f,m}}\left(\mathbf{s}_{m}\right).
\end{equation}
Following \cite{nazer2011compute}, we choose $\Lambda_{f,m}$ as
the finest lattice in $\left\{ \Lambda_{c,l},\textrm{for }l\textrm{ with }a_{ml}\neq0\right\} $,
i.e. 
\begin{equation}
\Lambda_{f,m}\triangleq\textrm{fine}\left\{ \Lambda_{c,l},\textrm{for }l\textrm{ with }a_{ml}\neq0\right\} \label{eq:Lambda_fm}
\end{equation}
Note that $\bm{\delta}_{m}$ in \eqref{eq:omega-def} is an integer
linear combination of $\mathbf{t}_{1},\cdots,\mathbf{t}_{L}$, together
with some residual dithering signals. This implies that not only the
relays but also the destination is required to have the knowledge
of $\mathbf{d}_{l}$ for dither cancellation.

We now determine the rate constraint to ensure the success of computation
at relay $m$. In \eqref{eq:s_m-expression}, $\tilde{\mathbf{z}}_{m}\triangleq\sum_{l=1}^{L}\theta_{ml}\mathbf{x}_{l}+\alpha_{m}\mathbf{z}_{m}$
is the equivalent noise. An error in computing $\hat{\bm{\delta}}_{m}$
occurs when the equivalent noise $\tilde{\mathbf{z}}_{m}$ lies outside
the fundamental Voronoi region of $\Lambda_{d,m}$. This error probability
goes to zero, i.e. 
\begin{equation}
\lim_{n\to\infty}\Pr\left\{ \hat{\bm{\delta}}_{m}\neq\bm{\delta}_{m}\right\} =0\label{eq:error-prob-goto-0}
\end{equation}
provided 
\begin{equation}
V_{c,l}=\textrm{Vol}\left(\mathcal{V}_{c,l}\right)>\left(2\pi e\max_{m:a_{ml}\neq0}\tau_{m}\right)^{n/2}\label{eq:V_fl_constraint}
\end{equation}
where $\tau_{m}=\alpha_{m}^{2}+\sum_{l=1}^{L}\left(\alpha_{m}h_{ml}-a_{ml}\right)^{2}p_{l}$
is the power of $\tilde{\mathbf{z}}_{m}$, and $\mathbf{a}_{m}=\left[a_{m1},\cdots,a_{mL}\right]^{T}$. 

Let $\mathbf{P}=\textrm{diag}\left(p_{1},p_{2},\cdots,p_{L}\right)$,
and $\mathbf{P}^{\frac{1}{2}}=\textrm{diag}\left(\sqrt{p_{1}},\sqrt{p_{2}},\cdots,\sqrt{p_{L}}\right)$.
By \eqref{eq:rate_nested_lattice_code}, \eqref{eq:lattice_average_power},
and \eqref{eq:V_fl_constraint}, the rate of the $l$-th source is
given by 
\begin{eqnarray}
r_{l} & = & \frac{1}{n}\log\frac{\textrm{Vol}\left(\mathcal{V}_{s,l}\right)}{\textrm{Vol}\left(\mathcal{V}_{c,l}\right)}\nonumber \\
 & < & \frac{1}{2}\log\negmedspace\left(\negmedspace\frac{p_{l}}{\underset{m:a_{ml}\neq0}{\max}\negmedspace\left[\alpha_{m}^{2}\negmedspace+\negmedspace\sum_{\rho=1}^{L}\left(\alpha_{m}h_{m\rho}\negmedspace-\negmedspace a_{m\rho}\right)^{2}p_{j}\right]}\negmedspace\right)\nonumber \\
 & = & \frac{1}{2}\log\left(\negmedspace\frac{p_{l}}{\underset{m:a_{ml}\neq0}{\max}\negmedspace\left[\alpha_{m}^{2}\negmedspace+\negmedspace\left\Vert \mathbf{P}^{\frac{1}{2}}\left(\alpha_{m}\mathbf{h}_{m}\negmedspace-\negmedspace\mathbf{a}_{m}\right)\right\Vert ^{2}\right]}\negmedspace\right).\label{eq:rate_tuple}
\end{eqnarray}
Note that the rate expression in \eqref{eq:rate_tuple} reduces to
the rate in Theorem 5 of \cite{nazer2011compute} by letting $p_{1}=p_{2}=\cdots=p_{L}$.
We will show that, instead of fixing $p_{l}=P_{l}$, allowing $p_{l}<P_{l}$
leads to a considerable performance gain.

We now optimize $\left\{ \alpha_{m}\right\} $ to obtain better computation
rates. Denote 
\begin{equation}
\varphi_{m}\left(\alpha_{m}\right)=\alpha_{m}^{2}+\left\Vert \mathbf{P}^{\frac{1}{2}}\left(\alpha_{m}\mathbf{h}_{m}-\mathbf{a}_{m}\right)\right\Vert ^{2}.
\end{equation}
By letting $\frac{\partial}{\partial\alpha_{m}}\varphi_{m}\left(\alpha_{m}\right)=0$,
we obtain an MMSE coefficient as 
\begin{equation}
\alpha_{m}^{\textrm{opt}}=\frac{\mathbf{h}_{m}^{T}\mathbf{P}\mathbf{a}_{m}}{1+\left\Vert \mathbf{P}^{\frac{1}{2}}\mathbf{h}_{m}\right\Vert ^{2}}.
\end{equation}
Substituting $\alpha_{m}^{\textrm{opt}}$ in \eqref{eq:rate_tuple},
we obtain \begin{subequations}\label{eq:first-hop-rate} 
\begin{align}
 & r_{l}<\frac{1}{2}\log^{+}\left(\frac{p_{l}}{\underset{m:a_{ml}\neq0}{\max}\varphi_{m}\left(\alpha_{m}^{\textrm{opt}}\right)}\right)\label{eq:first-hop-rate-constraint-phi}\\
= & \frac{1}{2}\log^{+}\left(\underset{m:a_{ml}\neq0}{\min}\frac{p_{l}}{\left\Vert \mathbf{P}^{\frac{1}{2}}\mathbf{a}_{m}\right\Vert ^{2}-\frac{\left(\mathbf{h}_{m}^{T}\mathbf{P}\mathbf{a}_{m}\right)^{2}}{1+\left\Vert \mathbf{P}^{\frac{1}{2}}\mathbf{h}_{m}\right\Vert ^{2}}}\right)\triangleq\tilde{r}_{l}\label{eq:first-hop-rate-constraint}
\end{align}
\end{subequations}
where $\left(x\right)^{+}\triangleq\max\left(x,0\right)$.
To summarize, a computation rate tuple $\left(r_{1},r_{2},\cdots,r_{L}\right)$
is achievable, i.e., \eqref{eq:error-prob-goto-0} is met, in the
first hop if $r_{l}<\tilde{r}_{l}$, for $l=1,\cdots L$. Note that
\eqref{eq:first-hop-rate} reduces to the rate expression in \cite[Theorem 2]{nazer2011compute}
by letting $P_{1}=\cdots=P_{L}=P$.

\section{Second Hop: Forwarding to Destination\label{sec:Forwarding}}

The preceding section is focused on the decoding operation at relays.
In what follows, we focus on how to forward the decoded combinations
$\hat{\mathbf{\bm{\delta}}}_{m}$ to the destination.

\subsection{Compressing at Relays}

As illustrated in Fig. \ref{fig:Compute-and-forward.}, after computing,
each relay $m$ compresses $\hat{\mathbf{\bm{\delta}}}_{m}$ and forwards
$\mathbf{x}_{m}^{\prime}\in\mathbb{R}^{n}$ to the destination. $\left\{ \hat{\mathbf{\bm{\delta}}}_{m}\right\} $
at different relays are correlated, as they are combinations of the
same set of source messages. Thus, the relay's re-encoding problem
is a distributed source coding problem. Forwarding $\left\{ \hat{\mathbf{\bm{\delta}}}_{m}\right\} $
directly at the relays in general leads to information redundancy
at the destination. 

We propose the following two operations for the relays to compress
$\left\{ \hat{\mathbf{\bm{\delta}}}_{m}\right\} $. First quantize
each $\hat{\mathbf{\bm{\delta}}}_{m}$ with lattice $\Lambda_{d,m}$,
i.e. 
\begin{equation}
\hat{\mathbf{\bm{\omega}}}_{m}=Q_{\Lambda_{d,m}}\left(\hat{\mathbf{\bm{\delta}}}_{m}\right).\label{eq:compress-func}
\end{equation}
Then, take modulo of each $\hat{\mathbf{\bm{\omega}}}_{m}$ over a
lattice $\Lambda_{e,m}$, i.e. 
\begin{equation}
\hat{\mathbf{v}}_{m}=\hat{\mathbf{\bm{\omega}}}_{m}\bmod\Lambda_{e,m}.\label{eq:general-v_m-omega_m-mod-lambda_em}
\end{equation}
The $\hat{\mathbf{v}}_{m}$ obtained in \eqref{eq:general-v_m-omega_m-mod-lambda_em}
is a lattice codeword in the $m$-th relay's equivalent codebook $\mathcal{C}_{m}^{\prime}$
generated by the lattice pair $\left(\Lambda_{d,m},\Lambda_{e,m}\right)$.
Thus, with the above quantization and modulo operations, the forwarding
rate of each relay $m$ is reduced to 
\begin{equation}
R_{m}\triangleq\left(\frac{1}{n}\log\frac{\textrm{Vol}\left(\mathcal{V}_{e,m}\right)}{\textrm{Vol}\left(\mathcal{V}_{d,m}\right)}\right)^{+},\label{eq:general-relay-forwarding-rate}
\end{equation}
where the forwarding rate $R_{m}$ is the rate of $\mathbf{x}_{m}^{\prime}$.
As illustrated in Fig. \ref{fig:Compute-and-forward.}, $\hat{\mathbf{v}}_{m}$
is then encoded as $\mathbf{x}_{m}^{\prime}=\mathcal{E}_{m}\left(\hat{\mathbf{v}}_{m}\right)$
and forwarded to the destination, where $\mathcal{E}_{m}\left(\cdot\right)$
is the re-encoding function of relay $m$.

\subsection{Decoding at the Destination}

The decoding at the destination consists of two steps: (i) to compute
$\hat{\mathbf{v}}_{m},m=1,\cdots,L$, from $\mathbf{y}^{\prime}$,
and (ii) to recover $\left\{ \mathbf{w}_{l}\right\} $ from $\left\{ \hat{\mathbf{v}}_{m}\right\} $.
Without loss of generality, let $\mathcal{R}$ be the capacity region
of the second-hop channel specified by $p\left(\mathbf{y}^{\prime}\vert\mathbf{x}_{1}^{\prime},\cdots,\mathbf{x}_{L}^{\prime}\right)$.
Then, in step (i), the destination can compute $\hat{\mathbf{v}}_{m}$
with a vanishing error probability, provided that the forwarding rate
tuple $\left(R_{1},\cdots,R_{L}\right)$ satisfies 
\begin{equation}
\left(R_{1},\cdots,R_{L}\right)\in\mathcal{R}.\label{eq:relay-forwarding-rate-constraint}
\end{equation}
The remaining issue is to recover $\left\{ \mathbf{w}_{l}\right\} $
from $\left\{ \hat{\mathbf{v}}_{m}\right\} $ in step (ii). We will
discuss the design of the modulo lattices $\left\{ \Lambda_{e,m}\right\} $
and the quantization lattices $\left\{ \Lambda_{d,m}\right\} $ to
guarantee the recoverability of $\left\{ \mathbf{w}_{l}\right\} $
in the subsequent sections.

\subsection{Further Discussions}

The rest of this paper is mainly focused on the design of $\left\{ \Lambda_{d,m}\right\} $
and $\left\{ \Lambda_{e,m}\right\} $. On one hand, it is desirable
to choose $\left\{ \Lambda_{d,m}\right\} $ as coarse as possible
and $\left\{ \Lambda_{e,m}\right\} $ as fine as possible, so as to
reduce the forwarding rates at relays. On the other hand, $\left\{ \Lambda_{d,m}\right\} $
cannot be too coarse and $\left\{ \Lambda_{e,m}\right\} $ cannot
be too fine, so as to ensure the recovery of the source messages at
destination. We will elaborate the design of $\left\{ \Lambda_{d,m}\right\} $
and $\left\{ \Lambda_{e,m}\right\} $ that ensures the recoverability
of $\left\{ \mathbf{w}_{l}\right\} $ from $\mathbf{y}^{\prime}$. 

For ease of discussion, we henceforth assume no error in relay computation
(i.e., $\hat{\mathbf{\bm{\delta}}}_{m}=\bm{\delta}_{m},\forall m$)
and destination computation (i.e. the destination can perfectly recover
$\left\{ \hat{\mathbf{v}}_{m}\right\} $ with $\mathbf{y}^{\prime}$).
Then, recovering $\left\{ \mathbf{w}_{l}\right\} $ from $\left\{ \mathbf{y}^{\prime}\right\} $
is equivalent to recovering $\left\{ \mathbf{w}_{l}\right\} $ from
$\left\{ \mathbf{v}_{m}\right\} $, where 
\begin{eqnarray}
\mathbf{v}_{m} & = & Q_{\Lambda_{d,m}}\left(\mathbf{\bm{\delta}}_{m}\right)\bmod\Lambda_{e,m}\label{eq:v_m-exp}
\end{eqnarray}
is the error-free version of $\hat{\mathbf{v}}_{m}$. Since $\mathbf{t}_{l}=\phi_{l}\left(\mathbf{w}_{l}\right)$
is an isomorphic mapping, recovering $\left\{ \mathbf{w}_{l}\right\} $
from $\left\{ \mathbf{v}_{m}\right\} $ is further equivalent to recovering
$\left\{ \mathbf{t}_{l}\right\} $ from $\left\{ \mathbf{v}_{m}\right\} $.

\section{Quantization at Relays\label{sec:Quantization-at-the-relays}}

In this section, we focus on the design of the quantization lattices
$\left\{ \Lambda_{d,m}\right\} $ to ensure the recoverability of
$\left\{ \mathbf{t}_{l}\right\} $ from $\left\{ \mathbf{v}_{m}\right\} $.
For convenience of discussion, we assume symmetric CF (SCF), i.e., 
all sources have the same power $p_{l}=p$
and thus the same shaping lattice $\Lambda_{s,l}=\Lambda_{s},\forall l$.
Then, the modulo lattices at relays can be trivially chosen as $\Lambda_{e,m}=\Lambda_{s},\forall m$.
Then, $\mathbf{v}_{m}$ in \eqref{eq:v_m-exp} becomes 
\begin{eqnarray}
\mathbf{v}_{m} & = & \negmedspace Q_{\Lambda_{d,m}}\left(\mathbf{\bm{\delta}}_{m}\right)\bmod\Lambda_{s}\nonumber \\
 & = & \negmedspace Q_{\Lambda_{d,m}}\negmedspace\left(\negthinspace\sum_{l=1}^{L}\negmedspace a_{ml}\left(\mathbf{t}_{l}\negthinspace-\negthinspace Q_{\Lambda_{s}}\negmedspace\left(\mathbf{t}_{l}\negthinspace-\negthinspace\mathbf{d}_{l}\right)\right)\negthinspace\right)\negmedspace\bmod\negmedspace\Lambda_{s}.\label{eq:v_m-exp-only-quan}
\end{eqnarray}

\subsection{Asymmetric Quantization Approach}

We now consider the design of the quantization lattices $\left\{ \Lambda_{d,m}\right\} $.
We choose $\Lambda_{d,m},m=1,\cdots,L$, to be a permutation of the
coding lattices $\Lambda_{c,l},l=1,\cdots,L$, i.e. 
\begin{equation}
\Lambda_{d,m}=\Lambda_{A,\pi_{d}\left(m\right)},m=1,\cdots,L\label{eq:asymmetric-quantization}
\end{equation}
where $\pi_{d}\left(\cdot\right)$ is a permutation function of $\left\{ 1,\cdots,L\right\} $.

The reason for the above choice of quantization lattices is explained
as follows. From \eqref{eq:rate_nested_lattice_code}, \eqref{eq:general-relay-forwarding-rate},
and \eqref{eq:asymmetric-quantization}, the forwarding rate of the
$m$-th relay is 
\begin{equation}
R_{m}=r_{\pi_{c}^{-1}\left(\pi_{d}\left(m\right)\right)}.\label{eq:asymmetric-relay-forwarding-rewritten-quan}
\end{equation}
As $\pi_{c}^{-1}\left(\pi_{d}\left(\cdot\right)\right)$ is a permutation,
we obtain $\sum_{m=1}^{L}R_{m}=\sum_{l=1}^{L}r_{l}$. To ensure that
the destination is able to recover all the source messages, the total
forwarding rate can not be less than $\sum_{l=1}^{L}r_{l}$. This
implies that we can not choose $\left\{ \Lambda_{d,m}\right\} $ finer
than \eqref{eq:asymmetric-quantization}. 

We say that $\pi_{d}\left(\cdot\right)$ is \textit{feasible} if the
destination can fully recover $\left\{ \mathbf{t}_{l}\right\} $ from
$\left\{ \mathbf{v}_{m}\right\} $. In general, the quantization $Q_{\Lambda_{d,m}}\left(\bm{\delta}_{m}\right)$
introduces information loss at relay $m$, and therefore $\pi_{d}\left(\cdot\right)$
in \eqref{eq:asymmetric-quantization} may be infeasible. We will
show that, in a multi-relay system, such information loss at a relay
does not necessarily translate to information loss at the destination.

\subsection{Heuristic Discussions}

We first introduce the following factorization of $\mathbf{t}_{l}$:
\begin{align}
 & \mathbf{t}_{l}=\mathbf{t}_{l}\bmod\Lambda_{s}\nonumber \\
= & \Bigl[Q_{\Lambda_{A,2}}\left(\mathbf{t}_{l}\right)+\underbrace{\mathbf{t}_{l}\bmod\Lambda_{A,2}}_{\mathbf{t}_{l,1}}\Bigl]\bmod\Lambda_{s}\nonumber \\
= & \Bigl[Q_{\Lambda_{A,3}}\left(Q_{\Lambda_{A,2}}\left(\mathbf{t}_{l}\right)\right)\negthinspace+\negthinspace\underbrace{Q_{\Lambda_{A,2}}\left(\mathbf{t}_{l}\right)\bmod\Lambda_{A,3}}_{\triangleq\mathbf{t}_{l,2}}+\mathbf{t}_{l,1}\Bigl]\bmod\Lambda_{s}\nonumber \\
= & \left[\sum_{\mu=1}^{L}\mathbf{t}_{l,\mu}\right]\bmod\Lambda_{s}\label{eq:lattice-codeword-split}
\end{align}
where $\mathbf{t}_{l,\mu}\in\Lambda_{A,\mu}\cap\mathcal{V}_{A,\mu+1}$,
for $\mu=1,\cdots,L$, (with $\mathcal{V}_{A,L+1}=\mathcal{V}_{s}$).
Each $\mathbf{t}_{l,\mu}$ is a representation of $\mathbf{t}_{l}$
in the lattice codebook $\Lambda_{A,\mu}\cap\mathcal{V}_{A,\mu+1}$.
Therefore, $\mathbf{t}_{l,\mu}=\mathbf{0}$ if the shaping lattice
$\Lambda_{A,\mu+1}$ is not coarser than the coding lattice of $\mathbf{t}_{l}$,
i.e. 
\begin{equation}
\mathbf{t}_{l}=\left[\sum_{\mu=\pi_{c}\left(l\right)}^{L}\mathbf{t}_{l,\mu}\right]\bmod\Lambda_{s}.\label{eq:lattice-codeword-split-simplified}
\end{equation}
Consider $\left\{ \mathbf{t}_{l,1},l=1,\cdots,L\right\} $, i.e.,
all the representations of $\left\{ \mathbf{t}_{l}\right\} $ in $\Lambda_{A,1}\cap\mathcal{V}_{A,2}$.
Only one of them, i.e., $\mathbf{t}_{\pi_{c}^{-1}\left(1\right),1}$,
is non-zero, since the finest coding lattice $\Lambda_{\negthinspace A,1}$
appears in $\left\{ \Lambda_{c,l}\right\} $ only once.

We now consider recovering $\mathbf{t}_{\pi_{c}^{-1}\left(1\right),1}$
from $\left\{ \mathbf{v}_{m}\right\} $. From \eqref{eq:asymmetric-quantization},
the only $\mathbf{v}_{m}$ that contains $\mathbf{t}_{\pi_{c}^{-1}\left(1\right),1}$
is $\mathbf{v}_{\pi_{d}^{-1}\left(1\right)}$. Compute 
\begin{align}
 & \mathbf{v}_{\pi_{d}^{-1}\left(1\right)}\bmod\Lambda_{A,2}\nonumber \\
= & Q_{\Lambda_{d,\pi_{d}^{-1}\left(1\right)}}\negthickspace\left(\negthinspace\sum_{l=1}^{L}\negmedspace a_{\pi_{d}^{-1}\negthinspace\left(1\right)l}\negmedspace\left(\mathbf{t}_{l}\negthinspace-\negthinspace Q_{\Lambda_{s}}\negmedspace\left(\mathbf{t}_{l}\negthinspace-\negthinspace\mathbf{d}_{l}\right)\right)\negmedspace\right)\negthickspace\bmod\negmedspace\Lambda_{s}\negmedspace\bmod\negmedspace\Lambda_{\negthinspace A,2}\nonumber \\
= & \left[\sum_{l=1}^{L}a_{\pi_{d}^{-1}\left(1\right)l}\mathbf{t}_{l}\right]\bmod\Lambda_{A,2}\nonumber \\
= & \left[\sum_{l=1}^{L}a_{\pi_{d}^{-1}\left(1\right)l}\mathbf{t}_{l,1}\right]\bmod\Lambda_{A,2}\nonumber \\
= & \left[a_{\pi_{d}^{-1}\left(1\right)\pi_{c}^{-1}\left(1\right)}\mathbf{t}_{\pi_{c}^{-1}\left(1\right),1}\right]\bmod\Lambda_{A,2}.\label{eq:v_m-finest-piece}
\end{align}
From \cite{nazer2011compute}, $\mathbf{t}_{\pi_{c}^{-1}\left(1\right),1}$
is recoverable from \eqref{eq:v_m-finest-piece} provided that $\kappa^{-1}\left(a_{\pi_{d}^{-1}\left(1\right)\pi_{c}^{-1}\left(1\right)}\right)\neq0$.
Then, the contribution of $\mathbf{t}_{\pi_{c}^{-1}\left(1\right),1}$
can be subtracted away from $\left\{ \mathbf{v}_{m}\right\} $.

We next consider recovering $\mathbf{t}_{\pi_{c}^{-1}\left(1\right),2}$
and $\mathbf{t}_{\pi_{c}^{-1}\left(2\right),2}$ which only appears
in $\mathbf{v}_{\pi_{d}^{-1}\left(1\right)}$ and $\mathbf{v}_{\pi_{d}^{-1}\left(2\right)}$.
By following similar steps in \eqref{eq:v_m-finest-piece}, we obtain
two linear equations of $\mathbf{t}_{\pi_{c}^{-1}\left(1\right),2}$
and $\mathbf{t}_{\pi_{c}^{-1}\left(2\right),2}$. Provided that this
linear system has a unique solution, we can recover $\mathbf{t}_{\pi_{c}^{-1}\left(1\right),2}$
and $\mathbf{t}_{\pi_{c}^{-1}\left(2\right),2}$. Continue this process
until all $\left\{ \mathbf{t}_{l,\mu}\right\} $ are recovered. Finally,
$\left\{ \mathbf{t}_{l}\right\} $ are reconstructed using \eqref{eq:lattice-codeword-split-simplified}.

\subsection{Successive Recovering Algorithm}

We now present a successive recovering algorithm by formalizing the
heuristic discussions in the preceding subsection. We first introduce
some definitions. Denote the coefficient matrix $\mathbf{A}=\left[\mathbf{a}_{1},\cdots,\mathbf{a}_{L}\right]^{T}\in\mathbb{Z}_{\gamma}^{L\times L}$.
We map matrix $\mathbf{A}$ to the corresponding matrix $\mathbf{Q}$
over $\mathbb{F}_{\gamma}$ using $\kappa\left(\cdot\right)$. Specifically,
define $\mathbf{Q}=\kappa^{-1}\left(\mathbf{A}\bmod\gamma\right)\in\mathbb{F}_{\gamma}^{L\times L}$,
where the $(i,j)$-th element of $\mathbf{Q}$, denoted by $q_{ij}$,
is given by $q_{ij}=\kappa^{-1}\left(a_{ij}\bmod\gamma\right)$. Define
the residual source set at the $j$-th iteration as 
\begin{equation}
\mathcal{S}^{\left\langle j\right\rangle }\triangleq\left\{ l\vert\pi_{c}\left(l\right)\leq j,l\in\left\{ 1,\cdots,L\right\} \right\} .\label{eq:S<j>}
\end{equation}
Define the residual relay set at the $j$-th iteration as 
\begin{equation}
\mathcal{T}^{\left\langle j\right\rangle }\triangleq\left\{ l\vert\pi_{d}\left(l\right)\leq j,l\in\left\{ 1,\cdots,L\right\} \right\} .\label{eq:T<j>}
\end{equation}
Define the \textit{residual coefficient matrix} $\mathbf{Q}^{\left\langle j\right\rangle }\in\mathbb{F}_{\gamma}^{j\times j}$
as the submatrix of $\mathbf{Q}$ with the rows indexed by $\mathcal{S}^{\left\langle j\right\rangle }$
and the columns indexed by $\mathcal{T}^{\left\langle j\right\rangle }$.
In the above, the superscript ``$^{\left\langle j\right\rangle }$''
represents the $j$-th iteration. Define the \textit{effective lattice
codebook for the $j$-th iteration} as $\mathcal{C}_{e}^{\left\langle j\right\rangle }$
generated by the lattice pair $\left(\Lambda_{A,j},\Lambda_{A,j+1}\right)$.
Define a mapping from $\mathbb{F}_{\gamma}^{k}$ to $\mathcal{C}_{e}^{\left\langle j\right\rangle }$
as 
\begin{equation}
\psi^{\left\langle j\right\rangle }\left(\mathbf{b}\right)\triangleq\left[\mathbf{B}\gamma^{-1}\kappa\left(\mathbf{G}\mathbf{b}\right)\right]\bmod\Lambda_{A,j+1}\label{eq:equivalent-lattice-encoding-quan}
\end{equation}
where 
\begin{equation}
\mathbf{b}=[\underbrace{0,\cdots,0}_{k_{A,j+1}},\underbrace{\tilde{\mathbf{\mathbf{b}}}^{T}}_{k_{A,j}-k_{A,j+1}},\underbrace{0,\cdots,0}_{k-k_{A,j}}]^{T}\in\mathbb{F}_{\gamma}^{k}.
\end{equation}
Denote by $\psi^{-\left\langle j\right\rangle }\left(\cdot\right)$
the inverse mapping of $\psi^{\left\langle j\right\rangle }\left(\cdot\right)$. 

We are now ready to present the \textit{Successive Recovering algorithm
for asymmetric Quantization operation} (referred to as the SRQ algorithm),
as presented in Algorithm \ref{alg:SRQ} below. 

\begin{algorithm}[h]
\begin{algorithmic}[1]

\Statex \textbf{Input}: $\left\{ \mathbf{v}_{m}\right\} _{m=1}^{L},\left\{ \Lambda_{c,l}\right\} _{l=1}^{L},\left\{ \Lambda_{d,m}\right\} _{m=1}^{L},\Lambda_{s}$

\Statex \textbf{Output}: $\hat{\mathbf{t}}_{l},l=1,\cdots,L$

\State \label{state:Initialization-q}\textbf{Initialization}: $\mathbf{v}_{m}^{\left\langle 1\right\rangle }\gets\mathbf{v}_{m}$
for $m=1,\cdots,L$.

\State \textbf{for} $j=1$ \textbf{to} $L$:

\State \quad{}\label{state:U-hat-q}Form a matrix $\mathbf{U}^{\left\langle j\right\rangle }$
by stacking $\left(\psi^{-\left\langle j\right\rangle }\left(\mathbf{v}_{m}^{\left\langle j\right\rangle }\bmod\Lambda_{A,j+1}\right)\right)^{T},m\in\mathcal{T}^{\left\langle j\right\rangle }$,
row by row.

\State \quad{}\label{state:recover-q}Recovering: Compute $\hat{\mathbf{W}}^{\left\langle j\right\rangle }\gets\left(\mathbf{Q}^{\left\langle j\right\rangle }\right)^{-1}\mathbf{U}^{\left\langle j\right\rangle }$;
construct $\left\{ \hat{\mathbf{w}}_{l,j}\right\} $ by de-stacking
$\hat{\mathbf{W}}^{\left\langle j\right\rangle }$ row by row, and
compute $\hat{\mathbf{t}}_{l,j}\gets\psi^{\left\langle j\right\rangle }\left(\hat{\mathbf{w}}_{l,j}\right)$,
for $l\in\mathcal{S}^{\left\langle j\right\rangle }$.

\State \quad{}\label{state:cancel-q}Cancellation: If $j<L$, compute
\begin{equation}
\mathbf{v}_{m}^{\left\langle j+1\right\rangle }\negmedspace\gets\negmedspace\left[\mathbf{v}_{m}-Q_{\Lambda_{d,m}}\negmedspace\left(\negthinspace\sum_{l\in\mathcal{S}^{\left\langle j\right\rangle }}a_{ml}\negmedspace\sum_{\mu=\pi_{c}\left(l\right)}^{j}\negmedspace\hat{\mathbf{t}}_{l,\mu}\negthinspace\right)\negthinspace\right]\negthickspace\bmod\Lambda_{s}\label{eq:SRQ-v_m}
\end{equation}
for $m\in\mathcal{T}^{\left\langle j+1\right\rangle }$.

\State\textbf{end for}

\State \label{state:combine-q}$\hat{\mathbf{t}}_{l}\gets\left[\sum_{\mu=\pi_{c}\left(j\right)}^{L}\hat{\mathbf{t}}_{l,\mu}\right]\bmod\Lambda_{s}$,
for $l=1,\cdots,L$. 

\end{algorithmic}

\caption{(SRQ algorithm)}
\label{alg:SRQ}
\end{algorithm}

We briefly explain Algorithm \ref{alg:SRQ} as follows. In Line \ref{state:U-hat-q},
$\psi^{-\left\langle j\right\rangle }$ maps a lattice point in $\mathcal{C}_{e}^{\langle j\rangle}$
to an integer vector in $\mathbb{F}_{\gamma}^{k}$; $\left\{ \Lambda_{A,j}\right\} $
and $\left\{ \Lambda_{d,m}\right\} $ are related by \eqref{eq:asymmetric-quantization};
$\Lambda_{A,L+1}=\Lambda_{s}$. In Line \ref{state:recover-q}, $\left(\mathbf{Q}^{\left\langle j\right\rangle }\right)^{-1}$
is the inverse of $\mathbf{Q}^{\left\langle j\right\rangle }$ in
$\mathbb{F}_{\gamma}^{j\times j}$. In Line \ref{state:cancel-q},
the contributions of $\left\{ \hat{\mathbf{t}}_{l,\mu},\forall l\in\mathcal{S}^{\left\langle j\right\rangle },\mu\leq j\right\} $
are cancelled from $\left\{ \mathbf{v}_{m}\right\} $. 

A sufficient condition to ensure the success of Algorithm \ref{alg:SRQ}
(i.e., $\hat{\mathbf{t}}_{l}=\mathbf{t}_{l},l=1,\cdots,L$) is presented
below. 

\begin{lem}
\label{lem:quan-lattice-decodability}Assume that $\mathbf{Q}^{\left\langle j\right\rangle },j=1,\cdots,L$,
are of full rank over $\mathbb{F}_{\gamma}$. Then the output of Algorithm
\ref{alg:SRQ} satisfies $\hat{\mathbf{t}}_{l}=\mathbf{t}_{l},l=1,\cdots,L$. 
\end{lem}

\begin{IEEEproof}
The proof is given in Appendix \ref{sec:Proof-of-SRQ}. 
\end{IEEEproof}

The following lemma states the existence of at least one feasible
$\pi_{d}\left(\cdot\right)$, i.e., all $\mathbf{Q}^{\left\langle j\right\rangle }$
are of full rank such that we can recover source messages correctly
by Lemma \ref{lem:quan-lattice-decodability}. 

\begin{lem}
\label{lem:pi_d-design}For given $\pi_{c}\left(\cdot\right)$, if
$\mathbf{Q}$ is of full rank over $\mathbb{F}_{\gamma}$, then there
exists a mapping $\pi_{d}\left(\cdot\right)$ satisfying \eqref{eq:asymmetric-quantization}
such that every residual coefficient matrix $\mathbf{Q}^{\left\langle j\right\rangle }$
is of full rank over $\mathbb{F}_{\gamma}$, $\forall j$. 
\end{lem}

\begin{IEEEproof}
The proof is given in Appendix \ref{sec:proof-pi_d-design}. 
\end{IEEEproof}

Note that feasible $\pi_{d}\left(\cdot\right)$ is in general not
unique. Thus, we need to search over all feasible $\pi_{d}\left(\cdot\right)$
in performance optimization, as elaborated later in Section \ref{sec:Performance-Optimization}.

\section{Modulo Operation at Relays\label{sec:Modulo-Operation-at-the-relays}}

The previous section is devoted to the design of relay's quantization
operations when the sources use asymmetric coding lattices. In this
section, we focus on the design of the modulo lattices $\left\{ \Lambda_{e,m}\right\} $
to reduce the forwarding rates. As analogous to the treatment in the
preceding section, we trivially choose the quantization lattices as
$\Lambda_{d,m}=\Lambda_{f,m},\forall m$. Then, $\mathbf{v}_{m}$
in \eqref{eq:v_m-exp} becomes 
\begin{eqnarray}
\mathbf{v}_{m} & = & \mathbf{\bm{\delta}}_{m}\bmod\Lambda_{e,m}\nonumber \\
 & = & \left[\sum_{l=1}^{L}\negmedspace a_{ml}\negthinspace\left(\mathbf{t}_{l}-Q_{\Lambda_{s,l}}\left(\mathbf{t}_{l}-\mathbf{d}_{l}\right)\right)\negthinspace\right]\negmedspace\bmod\negthinspace\Lambda_{e,m}.\label{eq:v_m-exp-only-mod}
\end{eqnarray}

\subsection{Asymmetric Modulo Operation\label{sub:Asymmetric-Modulo-Operation}}

Symmetric modulo operations have been previously used in CF \cite{nazer2011compute,ntranos2013usc}
to reduce the forwarding rates. In the symmetric modulo approach,
each relay $m$ takes modulo of $\hat{\mathbf{\bm{\delta}}}_{m}$
over the coarsest shaping lattice $\Lambda_{B,L}$, i.e., $\Lambda_{e,m}=\Lambda_{B,L}$,
for $m=1,\cdots,L$. Then the forwarding rate at relay $m$ is given
by 
\begin{equation}
R_{m}=\left(\frac{1}{n}\log\frac{\textrm{Vol}\left(\mathcal{V}_{B,L}\right)}{\textrm{Vol}\left(\mathcal{V}_{f,m}\right)}\right)^{+},\label{eq:symmetric-forwarding-rate}
\end{equation}
where $\mathcal{V}_{f,m}$ is the Voronoi region of $\Lambda_{f,m}$
defined in \eqref{eq:Lambda_fm}. Such a rate of $R_{m}$ may easily
exceed the maximum of the source rates, which implies information
redundancy. Therefore, the symmetric modulo approach is generally
far from optimal. 

To further reduce the forwarding rates, we propose an asymmetric modulo
approach to take modulo over different lattices at different relays.
Specifically, we assume that each relay $m$ takes modulo of $\hat{\mathbf{\bm{\delta}}}_{m}$
over $\Lambda_{e,m}$, with $\Lambda_{e,m},m=1,\cdots,L$, being a
permutation of the shaping lattices $\Lambda_{s,l},l=1,\cdots,L$,
i.e., 
\begin{equation}
\Lambda_{e,m}=\Lambda_{B,\pi_{e}\left(m\right)},m=1,\cdots,L\label{eq:asymmetric-modulo}
\end{equation}
where $\pi_{e}\left(\cdot\right)$ is a permutation function of $\left\{ 1,\cdots,L\right\} $.
Then we reduce the forwarding rate to 
\begin{equation}
R_{m}=\left(\frac{1}{n}\log\frac{\textrm{Vol}\left(\mathcal{V}_{B,\pi_{e}\left(m\right)}\right)}{\textrm{Vol}\left(\mathcal{V}_{f,m}\right)}\right)^{+}.\label{eq:asymmetric-forwarding-rate-asym-mod}
\end{equation}

For a random choice of $\pi_{e}\left(\cdot\right)$ in \eqref{eq:asymmetric-modulo},
the destination may be unable to recover $\left\{ \mathbf{t}_{l}\right\} $
from $\left\{ \mathbf{v}_{m}\right\} $. We say that $\pi_{e}\left(\cdot\right)$
is \textit{feasible} if the destination can correctly recover $\left\{ \mathbf{t}_{l}\right\} $
upon receiving $\left\{ \mathbf{v}_{m}\right\} $.

\subsection{Heuristic Discussions}

To recover $\left\{ \mathbf{t}_{l}\right\} $ from $\left\{ \mathbf{v}_{m}\right\} $,
the main idea is to convert the asymmetric system (with $\left\{ \mathbf{v}_{m}\right\} $
in \eqref{eq:general-v_m-omega_m-mod-lambda_em} defined using different
shaping lattices) into a series of symmetric ones (each with a common
shaping lattice). We start with the following observation on $\mathbf{v}_{\pi_{e}^{-1}\left(1\right)}$:
\begin{subequations}
\label{eq:modulo-fine}
\begin{align}
 & \mathbf{v}_{\pi_{e}^{-1}\left(1\right)}\bmod\Lambda_{B,1}\\
= & \left[\sum_{l=1}^{L}\negmedspace a_{ml}\negmedspace\left(\mathbf{t}_{l}-Q_{\Lambda_{s,l}}\negmedspace\left(\mathbf{t}_{l}-\mathbf{d}_{l}\right)\right)\right]\negmedspace\bmod\negmedspace\Lambda_{e,m}\negmedspace\bmod\negmedspace\Lambda_{B,1}\label{eq:modulo-fine-b}\\
= & \left[\sum_{l=1}^{L}a_{ml}\left(\mathbf{t}_{l}-Q_{\Lambda_{s,l}}\left(\mathbf{t}_{l}-\mathbf{d}_{l}\right)\right)\right]\bmod\Lambda_{B,1}\label{eq:modulo-fine-c}\\
= & \left[\sum_{l=1}^{L}a_{ml}\mathbf{t}_{l}\right]\bmod\Lambda_{B,1}\\
= & \left[\sum_{l}^{L}a_{ml}\left(\mathbf{t}_{l}\bmod\Lambda_{B,1}\right)\right]\bmod\Lambda_{B,1}\label{eq:converted-symmetric-system}
\end{align}
\end{subequations}
where $\Lambda_{B,1}$ is the finest shaping lattice used in the system, 
\eqref{eq:modulo-fine-b} follows from \eqref{eq:v_m-exp-only-mod},
and \eqref{eq:modulo-fine-c} follows from \eqref{eq:nested-lattice-mod-relation}.
Note that $\left(\mathbf{t}_{l}\bmod\Lambda_{B,1}\right)$ in \eqref{eq:converted-symmetric-system}
is a lattice codeword in the fundamental Voronoi region of $\Lambda_{B,1}$.
Thus, \eqref{eq:converted-symmetric-system} represents an effective
symmetric system where $\left\{ \mathbf{t}_{l}\bmod\Lambda_{B,1}\right\} $
are treated as source messages defined over a common shaping lattice
$\Lambda_{B,1}$, and every relay $m$ takes modulo of its received
combination $\sum_{l}a_{ml}\left(\mathbf{t}_{l}\bmod\Lambda_{B,1}\right)$
over $\Lambda_{B,1}$. Considering all the relays, this is a system
of $L$ equations with $L$ unknowns. It can be shown that the system
has a unique solution provided that the coefficient matrix $\mathbf{A}=\left[\mathbf{a}_{1},\cdots,\mathbf{a}_{L}\right]^{T}$
is of full rank over $\mathbb{F}_{\gamma}^{L\times L}$ \cite{nazer2011compute}.
However, the recovered $\left(\mathbf{t}_{l}\bmod\Lambda_{B,1}\right),l=1,\cdots,L$,
are in general not equal to $\mathbf{t}_{l}$, except $\mathbf{t}_{l}\bmod\Lambda_{B,1}=\mathbf{t}_{l}$
for $l=\pi_{s}^{-1}\left(1\right)$. This exception is because $\mathbf{t}_{\pi_{s}^{-1}\left(1\right)}$
is a lattice codeword in the Voronoi region of $\Lambda_{B,1}$, and
thus $\mathbf{t}_{\pi_{s}^{-1}\left(1\right)}\bmod\Lambda_{B,1}=\mathbf{t}_{\pi_{s}^{-1}\left(1\right)}$.
To recover other codewords, we cancel the contribution of the correctly
recovered $\mathbf{t}_{\pi_{s}^{-1}\left(1\right)}$ from the combinations
$\left\{ \hat{\mathbf{v}}_{m}\right\} $, and discard $\hat{\mathbf{v}}_{\pi_{e}^{-1}\left(1\right)}$
(as it is not useful in recovering other lattice codewords). In this
way, we obtain a residual asymmetric system, where the $\pi_{s}^{-1}\left(1\right)$-th
source and the $\pi_{e}^{-1}\left(1\right)$-th relay are deleted.
Then, we can recover $\mathbf{t}_{\pi_{s}^{-1}\left(2\right)}$ in
a similar way, and so forth. Finally, we can recover all $\left\{ \mathbf{t}_{l}\right\} $.

\subsection{Successive Recovering Algorithm\label{sub:SRM-alg}}

To make the above heuristic idea concrete, we introduce the following
definitions. In the $i$-th iteration, the \textit{residual source
set} is defined as 
\begin{equation}
\mathcal{S}^{\left(i\right)}\triangleq\left\{ l\vert\pi_{s}\left(l\right)\geq i,l\in\left\{ 1,\cdots,L\right\} \right\} ;\label{eq:S(i)}
\end{equation}
the \textit{residual relay set} is defined as 
\begin{equation}
\mathcal{T}^{\left(i\right)}\triangleq\left\{ l\vert\pi_{e}\left(m\right)\geq i,m\in\left\{ 1,\cdots,L\right\} \right\} ;\label{eq:T(i)}
\end{equation}
the \textit{residual coefficient matrix} $\mathbf{Q}^{\left(i\right)}\in\mathbb{F}_{\gamma}^{\left(L-i+1\right)\times\left(L-i+1\right)}$
is the submatrix of $\mathbf{Q}$ with the rows indexed by $\mathcal{S}^{\left(i\right)}$
and the columns indexed by $\mathcal{T}^{\left(i\right)}$. In the
above, the superscript ``$^{\left(i\right)}$'' represents the $i$-th
iteration. Define the \textit{effective lattice codebook for the $i$-th
iteration} as $\mathcal{C}_{e}^{\left(i\right)}$, generated by the
lattice pair $\left(\Lambda_{A,1},\Lambda_{B,i}\right)$. Define a
mapping from $\mathbb{F}_{\gamma}^{k}$ to $\mathcal{C}_{e}^{\left(i\right)}$
as 
\begin{equation}
\psi^{\left(i\right)}\left(\mathbf{b}\right)\triangleq\left[\mathbf{B}\gamma^{-1}\kappa\left(\mathbf{G}\mathbf{b}\right)\right]\bmod\Lambda_{B,i}\label{eq:equivalent-lattice-encoding}
\end{equation}
where $\mathbf{b}\in\mathbb{F}_{\gamma}^{k}$, and the first $k_{B,i}$
bits of $\mathbf{b}$ are all zero. Denote by $\psi^{-\left(i\right)}\left(\cdot\right)$
the inverse mapping of $\psi^{\left(i\right)}\left(\cdot\right)$.

We are now ready to present the \textit{Successive Recovering algorithm
for asymmetric Modulo operation} (referred to as the SRM algorithm),
as detailed below. 

\begin{algorithm}[H]
\begin{algorithmic}[1]

\Statex \textbf{Input}: $\left\{ \mathbf{v}_{m}\right\} _{m=1}^{L}$,
$\left\{ \mathbf{d}_{l}\right\} _{l=1}^{L}$, $\left\{ \Lambda_{c,l}\right\} _{l=1}^{L}$,
$\left\{ \Lambda_{d,m}\right\} _{m=1}^{L}$, $\left\{ \Lambda_{s,l}\right\} _{l=1}^{L}$,
$\left\{ \Lambda_{e,m}\right\} _{m=1}^{L}$

\Statex \textbf{Output}: $\hat{\mathbf{t}}_{l},l=1,\cdots,L$

\State \label{state:Initialization}\textbf{Initialization}: $\mathbf{v}_{m}^{\left(1\right)}\gets\mathbf{v}_{m}$
for all $m=1,\cdots,L$. 

\State \textbf{for} $i=1$ \textbf{to} $L$:

\State \quad{}\label{state:U-hat}Form $\mathbf{U}^{\left(i\right)}$
by stacking $\left(\psi^{-\left(i\right)}\left(\mathbf{v}_{m}^{\left(i\right)}\bmod\Lambda_{B,i}\right)\right)^{T},m\in\mathcal{T}^{\left(i\right)}$,
in a row-by-row manner. 

\State \quad{}\label{state:recover}Recovering: Compute $\hat{\mathbf{W}}^{\left(i\right)}\gets\left(\mathbf{Q}^{\left(i\right)}\right)^{-1}\mathbf{U}^{\left(i\right)}$;
set $\hat{\mathbf{w}}_{\pi_{s}^{-1}\left(i\right)}^{T}$ as the corresponding
row of $\hat{\mathbf{W}}^{\left(i\right)}$, and compute $\hat{\mathbf{t}}_{\pi_{s}^{-1}\left(i\right)}\gets\psi^{\left(i\right)}\left(\hat{\mathbf{w}}_{\pi_{s}^{-1}\left(i\right)}\right)$,
for $l\in\mathcal{S}^{\left\langle j\right\rangle }$. 

\State \quad{}\label{state:construct-v_i+1}Cancellation: If $j<L$,
then for $m\in\mathcal{T}^{\left(i+1\right)}$, denote $\tilde{\mathbf{t}}_{\pi_{s}^{-1}\left(i\right)}=\hat{\mathbf{t}}_{\pi_{s}^{-1}\left(i\right)}-Q_{\Lambda_{s,\pi_{s}^{-1}\left(i\right)}}\left(\hat{\mathbf{t}}_{\pi_{s}^{-1}\left(i\right)}-\mathbf{d}_{\pi_{s}^{-1}\left(i\right)}\right)$;
calculate 
\begin{equation}
\mathbf{v}_{m}^{\left(i+1\right)}\gets\Bigg[\mathbf{v}_{m}^{\left(i\right)}-a_{m\pi_{s}^{-1}\left(i\right)}\left(\tilde{\mathbf{t}}_{\pi_{s}^{-1}\left(i\right)}\right)\Bigg]\bmod\Lambda_{e,m}.\label{eq:SRM-v_m}
\end{equation}

\State \textbf{end for}

\end{algorithmic}
\caption{(SRM algorithm)}
\label{alg:SRM}
\end{algorithm}

Algorithm \ref{alg:SRM} is briefly explained as follows. In Line
\ref{state:U-hat}, we construct a symmetric system based on the $\bmod$-$\Lambda_{B,i}$
operation. In Line \ref{state:recover}, we recover source codeword
$\hat{\mathbf{t}}_{\pi_{s}^{-1}\left(i\right)}$ from the $\pi_{s}^{-1}\left(i\right)$-th
source. In Line \ref{state:construct-v_i+1}, we cancel the contribution
of $\hat{\mathbf{t}}_{\pi_{s}^{-1}\left(i\right)}$ from $\left\{ \mathbf{v}_{m}^{\left(i\right)},m\in\mathcal{T}^{\left(i+1\right)}\right\} $
to obtain $\mathbf{v}_{m}^{\left(i+1\right)}$.

\begin{lem}
\label{lem:mod-lattice-decodability}Assume that $\mathbf{Q}^{\left(i\right)},i=1,\cdots,L$,
are of full rank over $\mathbb{F}_{\gamma}$. Then the output of Algorithm
\ref{alg:SRM} satisfies $\hat{\mathbf{t}}_{l}=\mathbf{t}_{l},l=1,\cdots,L$. 
\end{lem}

The proof of Lemma \ref{lem:mod-lattice-decodability} is given in
Appendix \ref{sec:Proof-of-SRM}. Note that in Lemma \ref{lem:mod-lattice-decodability},
for given $\mathbf{Q}$ and $\left\{ p_{l}\right\} $, $\left\{ \mathbf{Q}^{\left(i\right)}\right\} $
is a function of $\mathcal{T}^{\left(i\right)}$, and thus a function
of $\pi_{e}\left(\cdot\right)$. Therefore, Lemma \ref{lem:mod-lattice-decodability}
gives a sufficient condition for the feasibility of $\pi_{e}\left(\cdot\right)$.
The following lemma ensures the existence of a feasible $\pi_{e}\left(\cdot\right)$
that makes all $\mathbf{Q}^{\left(i\right)}$ full-rank. Similarly
to $\pi_{d}\left(\cdot\right)$, feasible $\pi_{e}\left(\cdot\right)$
is in general not unique. 

\begin{lem}
\label{lem:pi_e-design}For given $\pi_{s}\left(\cdot\right)$, if
$\mathbf{Q}$ is of full rank over $\mathbb{F}_{\gamma}$, then there
exists a mapping $\pi_{e}\left(\cdot\right)$ such that every residual
coefficient matrix $\mathbf{Q}^{\left(i\right)}$ is of full rank
over $\mathbb{F}_{\gamma}$.
\end{lem}

\begin{IEEEproof}
The proof is given in Appendix \ref{sec:proof-pi_e-design}. 
\end{IEEEproof}

\section{Asymmetric Quantization \& Modulo Operation\label{sec:Asymmetric-Quan-Mod}}

We are now ready to consider the joint design of quantization lattices
$\left\{ \Lambda_{d,m}\right\} $ and modulo lattices $\left\{ \Lambda_{e,m}\right\} $
by combining the results obtained in the preceding two sections.

\subsection{Heuristic Discussions}

To start with, we now assume a general setting: the sources generally
have different coding and shaping lattices; the relays perform both
asymmetric quantization and modulo operation respectively defined
by \eqref{eq:asymmetric-quantization} and \eqref{eq:asymmetric-modulo}.
Then, $\mathbf{v}_{m}$ in \eqref{eq:v_m-exp} becomes 
\begin{eqnarray}
\negthickspace\mathbf{v}_{m}\negthickspace & = & Q_{\Lambda_{d,m}}\left(\mathbf{\bm{\delta}}_{m}\right)\bmod\Lambda_{e,m}\nonumber \\
 & = & Q_{\Lambda_{d,m}}\negthickspace\left(\negmedspace\sum_{l=1}^{L}\negmedspace a_{ml}\negmedspace\left(\mathbf{t}_{l}\negmedspace-\negmedspace Q_{\Lambda_{s,l}}\negmedspace\left(\mathbf{t}_{l}\negmedspace-\negmedspace\mathbf{d}_{l}\right)\right)\negmedspace\right)\negthickspace\bmod\negmedspace\Lambda_{e,m}\negthinspace.\label{eq:v_m-exp-quan-mod}
\end{eqnarray}
With permutations $\pi_{d}\left(\cdot\right)$ and $\pi_{e}\left(\cdot\right)$
for optimization, the forwarding rate of the $m$-th relay is given
by 
\begin{equation}
R_{m}=\left(\frac{1}{n}\log\frac{\textrm{Vol}\left(\mathcal{V}_{B,\pi_{e}\left(m\right)}\right)}{\textrm{Vol}\left(\mathcal{V}_{A,\pi_{d}\left(m\right)}\right)}\right)^{+}.\label{eq:forwarding-rate-asym-mod-quan}
\end{equation}
As a chain of shaping lattices are involved, we generalize \eqref{eq:lattice-codeword-split}
as 
\begin{eqnarray}
\mathbf{t}_{l} & = & \left[\sum_{\mu=1}^{L+1}\mathbf{t}_{l,\mu}\right]\bmod\Lambda_{s,l}\label{eq:lattice-codeword-split-general}
\end{eqnarray}
where $\mathbf{t}_{l,\mu}\in\Lambda_{A,\mu}\cap\mathcal{V}_{A,\mu+1}$,
for $\mu=1,\cdots,L$, (with $\mathcal{V}_{A,L+1}=\mathcal{V}_{B,1}$),
and $\mathbf{t}_{l,L+1}\in\Lambda_{B,1}\cap\mathcal{V}_{s,l}$. 

Our goal is still to recover $\left\{ \mathbf{t}_{l}\right\} $ from
$\left\{ \mathbf{v}_{m}\right\} $. To this end, we first take modulo
of $\mathbf{v}_{m}$ on $\Lambda_{B,1}$, yielding \begin{subequations}
\begin{align}
 & \thickspace\mathbf{v}_{m}\bmod\Lambda_{B,1}\\
= & \thickspace Q_{\Lambda_{d,m}}\negthickspace\left(\sum_{l=1}^{L}a_{ml}\left(\mathbf{t}_{l}-Q_{\Lambda_{s,l}}\left(\mathbf{t}_{l}-\mathbf{d}_{l}\right)\right)\right)\bmod\Lambda_{B,1}\\
= & \thickspace Q_{\Lambda_{d,m}}\left(\sum_{l=1}^{L}a_{ml}\mathbf{t}_{l}\right)\bmod\Lambda_{B,1}\label{eq:v_m-mod-Lambda-B1-b}\\
= & \thickspace Q_{\Lambda_{d,m}}\left(\sum_{l=1}^{L}a_{ml}\left(\mathbf{t}_{l}\bmod\Lambda_{B,1}\right)\right)\bmod\Lambda_{B,1},\label{eq:v_m-mod-Lambda-B1-c}
\end{align}
\end{subequations}where \eqref{eq:v_m-mod-Lambda-B1-b} follows by
noting $\Lambda_{B,1}\supseteq\Lambda_{s,l}$. It is not difficult
to see that \eqref{eq:v_m-mod-Lambda-B1-c} is a special case of \eqref{eq:v_m-exp-only-quan}
by letting $\Lambda_{s}=\Lambda_{B,1}$ and $\mathbf{d}_{l}=0$ in
\eqref{eq:v_m-exp-only-quan}. Thus, we use the SRQ algorithm to obtain
$\mathbf{t}_{l}\bmod\Lambda_{B,1},l=1,\cdots,L$. Afterwards, we cancel
the contributions of $\left\{ \mathbf{t}_{l}\bmod\Lambda_{B,1}\right\} $
from $\mathbf{v}_{m}$. Then, the resulting system is defined based
on a common coding lattice $\Lambda_{B,1}$, different shaping lattices
$\left\{ \Lambda_{s,l}\right\} $, together with the codewords $\left\{ Q_{\Lambda_{B,1}}\left(\mathbf{t}_{l}\right)\right\} $.
Thus, we use the SRM algorithm to further recover $\left\{ Q_{\Lambda_{B,1}}\left(\mathbf{t}_{l}\right)\right\} $.
Finally, we reconstruct $\left\{ \mathbf{t}_{l}\right\} $ as 
\[
\mathbf{t}_{l}=\mathbf{t}_{l}\bmod\Lambda_{B,1}+Q_{\Lambda_{B,1}}\left(\mathbf{t}_{l}\right),l=1,\cdots,L.
\]
Note that the details of the above cancellation and recovery process
can be found in Appendix \ref{sec:Proof-of-SRMQ}.

\subsection{Successive Recovering Algorithm}

We now formally present the \textit{Successive Recovering algorithm
for asymmetric Modulo and Quantization} (referred to as the SRMQ algorithm),
as shown in Algorithm \ref{alg:SRMQ}. In Line \ref{state:v_m-quan}
of Algorithm \ref{alg:SRMQ}, $\mathbf{v}_{m}^{\textrm{quan}}$ is
a lattice codeword in the $m$-th relay's equivalent codebook $\mathcal{C}_{m}^{\textrm{quan}}$
generated by the lattice pair $\left(\Lambda_{B,1},\Lambda_{e,m}\right)$. 

\begin{algorithm}[h]
\begin{algorithmic}[1]

\Statex \textbf{Input}: $\left\{ \mathbf{v}_{m}\right\} _{m=1}^{L}$,
$\left\{ \mathbf{d}_{l}\right\} _{l=1}^{L}$, $\left\{ \Lambda_{c,l}\right\} _{l=1}^{L}$,
$\left\{ \Lambda_{d,m}\right\} _{m=1}^{L}$, $\left\{ \Lambda_{s,l}\right\} _{l=1}^{L}$,
$\left\{ \Lambda_{e,m}\right\} _{m=1}^{L}$

\Statex \textbf{Output}: $\hat{\mathbf{t}}_{l},l=1,\cdots,L$

\State \label{state:joint-SRQ}$\left(\hat{\mathbf{t}}_{1}^{\textrm{quan}},\cdots,\hat{\mathbf{t}}_{L}^{\textrm{quan}}\right)\gets\textrm{SRQ}\bigl(\left\{ \mathbf{v}_{m}\bmod\Lambda_{B,1}\right\} _{m=1}^{L},$
$\left\{ \Lambda_{c,l}\right\} _{l=1}^{L},\left\{ \Lambda_{d,m}\right\} _{m=1}^{L},\Lambda_{B,1}\bigl)$.

\State \label{state:v_m-quan}$\mathbf{v}_{m}^{\textrm{quan}}\gets\left[\mathbf{v}_{m}-Q_{\Lambda_{d,m}}\left(\sum_{l=1}^{L}a_{ml}\hat{\mathbf{t}}_{l}^{\textrm{quan}}\right)\right]\bmod\Lambda_{e,m}$,
for $m=1,\cdots,L$. 

\State \label{state:dither-quan}$\mathbf{d}_{l}^{\textrm{quan}}\gets\mathbf{d}_{l}-\hat{\mathbf{t}}_{l}^{\textrm{quan}}$,
for $l=1,\cdots,L$. 

\State $\left(\hat{\mathbf{t}}_{1}^{\textrm{mod}},\cdots,\hat{\mathbf{t}}_{L}^{\textrm{mod}}\right)\gets\textrm{SRM}(\left\{ \mathbf{v}_{m}^{\textrm{quan}}\right\} _{m=1}^{L}$,
$\left\{ \mathbf{d}_{l}^{\textrm{quan}}\right\} _{l=1}^{L}$, $\left\{ \Lambda_{B,1},\cdots,\Lambda_{B,1}\right\} $,
$\left\{ \Lambda_{B,1},\cdots,\Lambda_{B,1}\right\} $, $\left\{ \Lambda_{s,l}\right\} _{l=1}^{L}$,
$\left\{ \Lambda_{e,m}\right\} _{m=1}^{L}$. 

\State $\hat{\mathbf{t}}_{l}\gets\hat{\mathbf{t}}_{l}^{\textrm{quan}}+\hat{\mathbf{t}}_{l}^{\textrm{mod}}$,
for $l=1,\cdots,L$. 

\end{algorithmic}

\protect\protect\caption{\label{alg:SRMQ}(SRMQ algorithm)}
\end{algorithm}

\begin{lem}
\label{lem:quan-mod-lattice-decodability}Assume that $\mathbf{Q}^{\left(i\right)}$
and $\mathbf{Q}^{\left\langle j\right\rangle }$, $j=1,\cdots,L$,
are of full rank over $\mathbb{F}_{\gamma}$. Then the output of Algorithm
\ref{alg:SRMQ} satisfies $\hat{\mathbf{t}}_{l}=\mathbf{t}_{l},l=1,\cdots,L$. 
\end{lem}

\begin{IEEEproof}
The proof is given in Appendix \ref{sec:Proof-of-SRMQ}.
\end{IEEEproof}

\subsection{Achievable Rates of the Overall Scheme}

We are now ready to present the achievable rates of the proposed scheme.
We have the following theorem.

\begin{thm}
\label{thm:achievable-rate-tuple-quan-mod} For given $\mathbf{H}$,
$\mathbf{A}$, and $\mathcal{R}$, a transmission rate tuple $\left(r_{1},r_{2},\cdots,r_{L}\right)$
is achievable if there exist $\pi_{d}\left(\cdot\right)$ and $\pi_{e}\left(\cdot\right)$,
such that the following conditions are met:\end{thm}
\begin{enumerate}
\item \label{enu:thm-const-pow-quan-mod}power constraints in \eqref{eq:lattice_average_power}:
$p_{l}\leq P_{l},\forall l$, 
\item \label{enu:thm:const-compute-quan-mod}computation constraints in
\eqref{eq:first-hop-rate-constraint}: $r_{l}<\tilde{r}_{l},\forall l$, 
\item \label{enu:thm:const-forward-quan-mod}forwarding rate constraints
in \eqref{eq:relay-forwarding-rate-constraint}: $\left(R_{1},\cdots,R_{L}\right)\in\mathcal{R}$,
where $R_{m},m=1,\cdots,L$, are given by 
\begin{equation}
R_{m}=\left(r_{\pi_{c}^{-1}\left(\pi_{d}\left(m\right)\right)}\negmedspace+\negmedspace\frac{1}{2}\log\negmedspace\left(\frac{p_{\pi_{s}^{-1}\left(\pi_{e}\left(m\right)\right)}}{p_{\pi_{c}^{-1}\left(\pi_{d}\left(m\right)\right)}}\right)\right)^{+}.\label{eq:asymmetric-relay-forwarding-rewritten-quan-mod}
\end{equation}
\item \label{enu:thm:const-recover-quan-mod}recovery constraints: all $\mathbf{Q}^{\left(i\right)}$
and $\mathbf{Q}^{\left\langle j\right\rangle }$ are of full rank
over $\mathbb{F}_{\gamma}$.
\end{enumerate}

\begin{IEEEproof}
Condition \ref{enu:thm-const-pow-quan-mod} is the power constraint
in \eqref{eq:lattice_average_power}. Condition \ref{enu:thm:const-compute-quan-mod}
ensures a vanishing computing error at every relay; see \eqref{eq:first-hop-rate}.
Also, $R_{m}$ in \eqref{eq:asymmetric-relay-forwarding-rewritten-quan-mod}
is obtained by noting \eqref{eq:good-for-MSE}, \eqref{eq:rate_nested_lattice_code},
\eqref{eq:lattice_average_power}, \eqref{eq:asymmetric-modulo},
\eqref{eq:asymmetric-quantization}, and \eqref{eq:forwarding-rate-asym-mod-quan}.
Finally, Condition \ref{enu:thm:const-recover-quan-mod} ensures the
recoverability of the source messages using the SRMQ algorithm as
specified in Lemma \ref{lem:quan-mod-lattice-decodability}. Therefore,
$\left(r_{1},r_{2},\cdots,r_{L}\right)$ is achievable under Conditions
\ref{enu:thm-const-pow-quan-mod} to \ref{enu:thm:const-recover-quan-mod}.\end{IEEEproof}

The following theorem ensures the existence of feasible $\pi_{d}\left(\cdot\right)$ 
and $\pi_{e}\left(\cdot\right)$ that make all 
$\mathbf{Q}^{\left\langle j\right\rangle }$ and $\mathbf{Q}^{\left(i\right)}$ full-rank. 

\begin{thm}
\label{thm:pi_d_e-design}For given $\pi_{c}\left(\cdot\right)$ and
$\pi_{s}\left(\cdot\right)$, if $\mathbf{Q}$ is of full rank over
$\mathbb{F}_{\gamma}$, then there exists mappings $\pi_{d}\left(\cdot\right)$
and $\pi_{e}\left(\cdot\right)$ such that all residual coefficient
matrices $\mathbf{Q}^{\left\langle j\right\rangle }$ and $\mathbf{Q}^{\left(i\right)}$
are of full rank over $\mathbb{F}_{\gamma}$. 
\end{thm}

\begin{IEEEproof}
Lemma \ref{lem:pi_d-design} and Lemma \ref{lem:pi_e-design} ensure
the existence of feasible $\pi_{d}\left(\cdot\right)$ and $\pi_{e}\left(\cdot\right)$
such that $\mathbf{Q}^{\left\langle j\right\rangle }$ and $\mathbf{Q}^{\left(i\right)}$,
$\forall i,j$, are of full rank.
\end{IEEEproof}

\section{Sum-Rate Maximization\label{sec:Performance-Optimization}}

In this section, we consider optimizing the achievable sum rate for
the proposed CCF scheme. A centralized node is assumed to acquire
all the knowledge of $\mathbf{H}$, $\mathcal{R}$, and $\left\{ P_{l}\right\} $.
This centralized node informs each source $l$ of lattice pair $\left(\Lambda_{c,l},\Lambda_{s,l}\right)$,
and each relay $m$ of quantization lattice $\Lambda_{d,m}$ and modulo
lattice $\Lambda_{e,m}$.

\subsection{Problem Formulation}

Based on Theorem \ref{thm:achievable-rate-tuple-quan-mod}, we formulate
the sum-rate maximization problem for the SRMQ algorithm as follows: 
\begin{subequations}\label{eq:opt-equations}
\begin{align}
\max_{\substack{\mathbf{A},\left\{ p_{l}\right\} ,\left\{ r_{l}\right\} \\
\pi_{d}\left(\cdot\right),\pi_{e}\left(\cdot\right)
}
} &  &  & \sum_{l=1}^{L}r_{l}\label{eq:opt-object}\\
\textrm{s.t.} &  &  & p_{l}\leq P_{l},\forall l,\\
 &  &  & r_{l}\negthinspace<\negthinspace\frac{1}{2}\negmedspace\log^{\negthinspace+}\negthickspace\left(\negthickspace\underset{m:a_{ml}\neq0}{\min}\frac{p_{l}}{\bigl\Vert\negthinspace\mathbf{P}^{\frac{1}{2}}\mathbf{a}_{m}\negthinspace\bigr\Vert{}^{\negmedspace2}-\negmedspace\frac{\left(\mathbf{h}_{m}^{T}\mathbf{P}\mathbf{a}_{m}\right){}^{2}}{1\negthinspace+\negthinspace\left\Vert \negthinspace\mathbf{P}^{\negthinspace\frac{1}{2}}\negthinspace\mathbf{h}_{m}\negthinspace\right\Vert {}^{2}}}\negthickspace\right)\negthickspace,\negmedspace\forall l,\label{eq:opt-computation-constraint}\\
 &  &  & \left(R_{1},\cdots,R_{L}\right)\in\mathcal{R},\label{eq:opt-second-hop-constraint}\\
 &  &  & R_{m}=r_{\pi_{c}^{-1}\left(\pi_{d}\left(m\right)\right)}\negmedspace+\negmedspace\frac{1}{2}\negmedspace\log\negmedspace\left(\negthinspace\frac{p_{\pi_{s}^{-1}\left(\pi_{e}\left(m\right)\right)}}{p_{\pi_{c}^{-1}\left(\pi_{d}\left(m\right)\right)}}\negthinspace\right)\negmedspace,\label{eq:opt-R_m-exp}\\
 &  &  & \textrm{rank}\left(\mathbf{Q}^{\left(i\right)}\right)=L-i+1,i=1,\cdots,L,\label{eq:opt-q-mod-constraint}\\
 &  &  & \textrm{rank}\left(\mathbf{Q}^{\left\langle j\right\rangle }\right)=j,j=1,\cdots,L.\label{eq:opt-q-quan-constraint}
\end{align}
\end{subequations}
Note that if the SRM algorithm is used for message
recovery, then $R_{m}$ in \eqref{eq:opt-R_m-exp} should be replaced
by 
\begin{equation}
R_{m}=\underset{l:a_{ml}\neq0}{\max}\left(r_{l}+\frac{1}{2}\log\left(\frac{p_{\pi_{s}^{-1}\left(\pi_{e}\left(m\right)\right)}}{p_{l}}\right)\right).\label{eq:asymmetric-relay-forwarding-rewritten}
\end{equation}
Similarly, if the SRQ algorithm is used for recovery, then \eqref{eq:opt-R_m-exp}
should be replaced by \eqref{eq:asymmetric-relay-forwarding-rewritten-quan}.

\subsection{Approximate Solution}

The problem in \eqref{eq:opt-equations} is an NP-hard mixed integer
program. Here we present a suboptimal solution as follows. For given
$\left\{ p_{l}\right\} $, we apply the LLL algorithm \cite{osmane2011compute,sakzad2012integer}
to determine the integer matrix $\mathbf{A}$. For given $\mathbf{A}$
and $\left\{ p_{l}\right\} $, we search over all permutations $\pi_{d}\left(\cdot\right)$
and all permutations $\pi_{e}\left(\cdot\right)$ to maximize \eqref{eq:opt-object}.
This can be done by noting that, given $\mathbf{A}$, $\left\{ p_{l}\right\} $,
$\pi_{d}\left(\cdot\right)$, and $\pi_{e}\left(\cdot\right)$, the
problem in \eqref{eq:opt-equations} is convex (as the capacity region
$\mathcal{R}$ is always a convex set). Also not that the feasible
set of $\pi_{d}\left(\cdot\right)$ and $\pi_{e}\left(\cdot\right)$
is not empty, as ensured by Theorem \ref{thm:pi_d_e-design}. The
above discussions are based on fixed $\left\{ p_{l}\right\} $. Finally,
we need to maximize \eqref{eq:opt-object} by exhaustively searching
over the quantized values of $\left\{ p_{l}\right\} $. 

We now give more details on determining $\mathbf{A}$ using the LLL
algorithm. We determine $\mathbf{A}$ in a column-by-column manner
from $\mathbf{a}_{1}$ to $\mathbf{a}_{L}$. For any step $m$, we
first note that maximizing the right hand side of \eqref{eq:opt-computation-constraint}
is equivalent to minimizing $\varphi_{m}\left(\alpha_{m}^{\textrm{opt}}\right)$
in \eqref{eq:first-hop-rate-constraint-phi}. We rewrite $\varphi_{m}\left(\alpha_{m}^{\textrm{opt}}\right)$
as 
\begin{equation}
\varphi_{m}\negmedspace\left(\alpha_{m}^{\textrm{opt}}\right)\negmedspace=\negmedspace\mathbf{a}_{m}^{T}\mathbf{P}\left(\negmedspace\mathbf{I}_{L}\negmedspace-\negmedspace\frac{\mathbf{P}^{T}\mathbf{h}_{m}\mathbf{h}_{m}^{T}\mathbf{P}}{1\negmedspace+\negmedspace\left\Vert \mathbf{P}\mathbf{h}_{m}\right\Vert ^{2}}\negmedspace\right)\negmedspace\mathbf{P}\mathbf{a}_{m}\negmedspace\triangleq\negmedspace\mathbf{a}_{m}^{T}\mathbf{D}_{m}\mathbf{a}_{m}\label{eq:phi_m-D_m}
\end{equation}
where 
\begin{equation}
\mathbf{D}_{m}\triangleq\mathbf{P}\left(\mathbf{I}_{L}-\frac{\mathbf{P}^{T}\mathbf{h}_{m}\mathbf{h}_{m}^{T}\mathbf{P}}{1+\left\Vert \mathbf{P}\mathbf{h}_{m}\right\Vert ^{2}}\right)\mathbf{P}
\end{equation}
is symmetric and positive definite. Cholesky decomposition of $\mathbf{D}_{m}$
gives $\mathbf{D}_{m}=\mathbf{L}_{m}\mathbf{L}_{m}^{T}$, and then
$\mathbf{D}_{m}$ is the Gram matrix for a lattice with generator
matrix $\mathbf{L}_{m}$. Apply the LLL algorithm to $\mathbf{L}_{m}$
to find the reduced matrix $\mathbf{L}_{m}^{\prime}$. Compute $\mathbf{A}_{m}=\mathbf{L}_{m}^{\prime}\mathbf{L}_{m}^{-1}$.
Then we choose $\mathbf{a}_{m}$ as the row of $\mathbf{A}_{m}$ with
the smallest norm that is linearly independent of $\mathbf{a}_{l}$,
$l=1,\cdots,m-1$. This ensures that the constructed integer matrix
$\mathbf{A}$ is of full rank over $\mathbb{F}_{\gamma}$. 

The above algorithm involves an exhaustive search over $\left\{ p_{l}\right\} $,
$\pi_{e}\left(\cdot\right)$ and $\pi_{d}\left(\cdot\right)$. Thus,
it is still computationally intensive, especially when the network
size is relatively large. The study of more efficient algorithms for
solving \eqref{eq:opt-equations} is out of the scope of this paper.

\subsection{Numerical Results}

To keep the computation complexity tractable, we set the second-hop
channel be parallel channels as 
\[
\mathbf{y}_{m}^{\prime}=g_{m}\mathbf{x}_{m}^{\prime}+\mathbf{z}_{m}^{\prime}
\]
where $g_{m}\in\mathbb{R}$ is the channel coefficient from relay
$m$ to the destination, $g_{m}\sim\mathcal{N}\left(0,1\right)$,
$\mathbf{x}_{m}^{\prime}\in\mathbb{R}^{n\times1}$ is the forwarded
signal at the $m$-th relay, $\frac{1}{n}\left\Vert \mathbf{x}_{m}^{\prime}\right\Vert ^{2}=P_{R,m}$,
and $\mathbf{z}_{m}^{\prime}\in\mathbb{R}^{n\times1}$ is i.i.d. Gaussian
noise, $\mathbf{z}_{m}^{\prime}\sim\mathcal{N}\left(\mathbf{0},\mathbf{I}_{n}\right)$.
So the rate region $\mathcal{R}$ is a hypercube in $\mathbb{R}^{L}$. 

In simulation, the following settings are employed: $L=2$; $P_{l}=P,\forall l$;
$N_{\textrm{brute}}=100$; $P_{R,m}=0.25P,\forall m$. Note that $N_{\textrm{brute}}$
represent the number of discretized power levels in exhaustively searching
each $p_{l}$.

\subsubsection{Comparison of Different Approaches in a Two-Hop System}

We now compare the following schemes: 
\begin{itemize}
\item SCF: the original CF in \cite{nazer2011compute};
\item SCF-Q: SCF with asymmetric quantization 
(using the SRQ algorithm for recovering);
\item ACF: ACF with symmetric modulo and conventional quantization;
\item ACF-M: ACF with asymmetric modulo and conventional quantization 
(using the SRM algorithm);
\item ACF-MQ: ACF with asymmetric modulo and asymmetric quantization 
(using the SRMQ algorithm).
\end{itemize}
\begin{figure}[h]
\centering
\includegraphics[width=0.7\columnwidth]{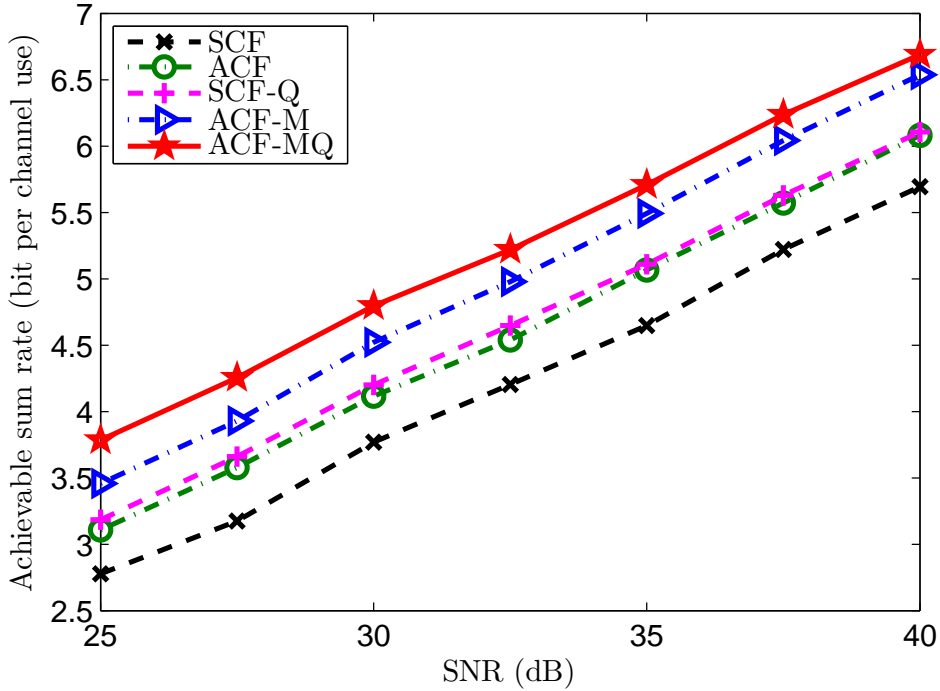}
\caption{Performance comparison in a two-hop system.}
\label{fig:CF-comparison-two-hop}
\end{figure}

The simulated sum-rate performances are compared in Fig. \ref{fig:CF-comparison-two-hop}.
We see that ACF outperforms SCF by about $2\textrm{ dB}$. Also, SCF-Q
outperforms SCF by about $2\textrm{ dB}$, due to the use of asymmetric
quantization; ACF-M outperforms ACF by about $2\textrm{ dB}$, due
to the use of asymmetric modulo operation; ACF-MQ outperforms ACF
by about $3\textrm{ dB}$, due to the use of both asymmetric modulo
operation and asymmetric quantization; ACF-MQ outperforms SCF by about
$5\textrm{ dB}$, thanks to the use of asymmetric shaping lattices,
asymmetric modulo operation, and asymmetric quantization. 
Fig. \ref{fig:CF-comparison-two-hop} demonstrates clearly 
the the performance advantage
of the proposed CCF scheme over the conventional CF scheme.

\subsubsection{Performance Comparison With Different Network Sizes}

\begin{figure}[tbh]
\centering
\includegraphics[width=0.7\columnwidth]{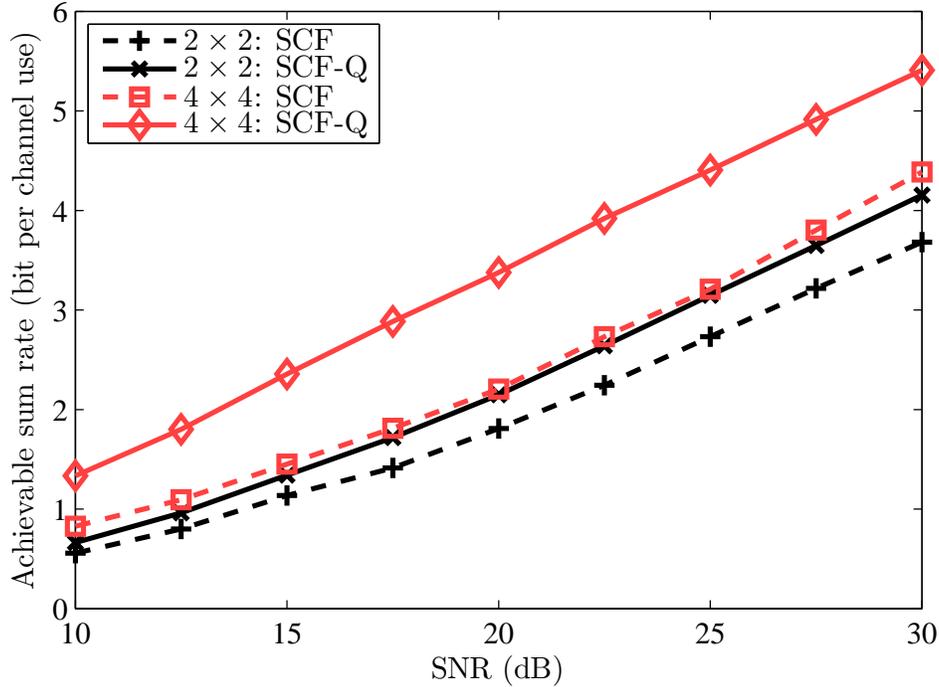}
\caption{Performance comparison with different network sizes.}
\label{fig:two-hop-size}
\end{figure}

We now provide the performance comparison with different network sizes.
Because of high computation complexity, 
we only simulate the performance of SCF and SCF-Q schemes. 
In Fig. \ref{fig:two-hop-size}, we see that, for a $2\times2$
network (with two sources and two relays), the power gain of SCF-Q
over SCF is about $2\textrm{ dB}$ at high SNR; for a $4\times4$
network, the corresponding power gain is about $6\textrm{ dB}$. We
see that the performance gain of SCF-Q increases significantly with
the network size.

It is highly desirable to understand how the 
performance of CCF improves with the network size. 
This requires the development of more efficient algorithms 
to solve \eqref{eq:opt-equations},
which is nevertheless an interesting topic for future research. 

\section{Conclusions\label{sec:Conclusions}}

In this paper, we proposed a novel relay strategy, named CCF, to exploit
the asymmetry naturally inherent in wireless networks. Compared with
conventional CF, CCF includes an extra compressing stage in between
the computing and forwarding stages. We proposed to use quantization
and modulo operation for compressing the message combinations computed
at relays, which reduces the information redundancy in the messages
forwarded by the relays and thereby improve the spectral efficiency
of the network. Particularly, we studied CCF design in a two-hop wireless
relaying network involving multiple sources, multiple relays, and
a single destination. We derived achievable rates of the scheme and
formulate a sum-rate maximization problem for performance optimization.
Numerical results were presented to demonstrate the significant performance
advantage of CCF over CF.

\appendices{}

\section{Proof of Lemma \ref{lem:quan-lattice-decodability}\label{sec:Proof-of-SRQ}}

We follow Algorithm \ref{alg:SRQ} step by step to prove that the
output of Algorithm \ref{alg:SRQ} satisfies $\hat{\mathbf{t}}_{l}=\mathbf{t}_{l},l=1,\cdots,L$.
Recall that Algorithm \ref{alg:SRQ} involves $L$ iterations. In
the $j$-th iteration, $\left\{ \hat{\mathbf{t}}_{l,j},l\in\mathcal{S}^{\left\langle j\right\rangle }\right\} $
is computed using $\left\{ \mathbf{v}_{m}^{\left\langle j\right\rangle },m\in\mathcal{T}^{\left\langle j\right\rangle }\right\} $,
where $\mathbf{v}_{m}^{\left\langle j\right\rangle }$ is set as $\mathbf{v}_{m}$
for $j=1$, and given in \eqref{eq:SRQ-v_m} for $j>1$.

We first prove that, in the $j$-th iteration, 
\begin{equation}
\mathbf{v}_{m}^{\left\langle j\right\rangle }=\left[\sum_{l=1}^{L}a_{ml}\sum_{\mu=j}^{L}\mathbf{t}_{l,\mu}\right]\bmod\Lambda_{s}\label{eq:quantized-v_m^i-after-split}
\end{equation}
for $m\in\mathcal{T}^{\left\langle j\right\rangle }$. We prove this
statement by induction. In initialization, we have $\mathbf{v}_{m}^{\left\langle 1\right\rangle }=\mathbf{v}_{m},m\in\mathcal{T}^{\left\langle 1\right\rangle }$.
For $j=1$, the residual relay set only contains one element, i.e.,
$\mathcal{T}^{\left\langle 1\right\rangle }=\left\{ \pi_{d}^{-1}\left(1\right)\right\} $,
and $\Lambda_{d,\pi_{d}^{-1}\left(1\right)}=\Lambda_{A,1}$ is the
finest lattice in $\left\{ \Lambda_{c,l}\right\} $. Then \begin{subequations}
\begin{align}
 & \medspace\mathbf{v}_{\pi_{d}^{-1}\left(1\right)}^{\left\langle 1\right\rangle }=\mathbf{v}_{\pi_{d}^{-1}\left(1\right)}\\
= & \medspace\left[\sum_{l=1}^{L}a_{\pi_{d}^{-1}\left(1\right)l}\left(\mathbf{t}_{l}-Q_{\Lambda_{s}}\left(\mathbf{t}_{l}-\mathbf{d}_{l}\right)\right)\right]\bmod\Lambda_{s}\label{eq:SRQ-recover-1-b}\\
= & \medspace\left[\sum_{l=1}^{L}a_{\pi_{d}^{-1}\left(1\right)l}\sum_{\mu=1}^{L}\mathbf{t}_{l,\mu}\right]\bmod\Lambda_{s}\label{eq:SRQ-recover-1-c}
\end{align}
\end{subequations}where \eqref{eq:SRQ-recover-1-b} follows from
\eqref{eq:v_m-exp-only-quan} and $\sum_{l=1}^{L}a_{\pi_{d}^{-1}\left(1\right)l}\left(\mathbf{t}_{l}-Q_{\Lambda_{s}}\left(\mathbf{t}_{l}-\mathbf{d}_{l}\right)\right)\in\Lambda_{A,1}$;
\eqref{eq:SRQ-recover-1-c} follows from \eqref{eq:lattice-codeword-split}.
Thus \eqref{eq:quantized-v_m^i-after-split} holds for $j=1$.

Now suppose that \eqref{eq:quantized-v_m^i-after-split} holds for
the $j$-th iteration. We show that $\hat{\mathbf{t}}_{l,j}=\mathbf{t}_{l,j},l\in\mathcal{S}^{\langle j\rangle}$
in the $j$-th iteration, and that \eqref{eq:quantized-v_m^i-after-split}
still holds for the $\left(j+1\right)$-th iteration.

In Line \ref{state:U-hat-q}, with $\left\{ \hat{\mathbf{v}}_{m}^{\left\langle j\right\rangle }\right\} $,
we obtain 
\begin{subequations}\label{eq:SRQ-recover-j}
\begin{eqnarray}
\hat{\mathbf{v}}_{m}^{\left\langle j\right\rangle }\negmedspace\bmod\negmedspace\Lambda_{A,j+1} & = & \left[\negthinspace\sum_{l=1}^{L}\negthinspace a_{ml}\negthinspace\sum_{\mu=j}^{L}\negthinspace\mathbf{t}_{l,\mu}\negthinspace\right]\negthickspace\bmod\negmedspace\Lambda_{A,j+1}\label{eq:SRQ-recover-j-a}\\
 & = & \left[\sum_{l=1}^{L}a_{ml}\mathbf{t}_{l,j}\right]\bmod\Lambda_{A,j+1}\label{eq:SRQ-recover-j-b}\\
 & = & \left[\sum_{l\in\mathcal{S}^{\left\langle j\right\rangle }}\negmedspace a_{ml}\mathbf{t}_{l,j}\right]\negthickspace\bmod\Lambda_{A,j+1}\label{eq:SRQ-recover-j-c}
\end{eqnarray}
\end{subequations}for all $m\in\mathcal{T}^{\left\langle j\right\rangle }$,
where \eqref{eq:SRQ-recover-j-a} holds since \eqref{eq:quantized-v_m^i-after-split}
is true for the $j$-th iteration; \eqref{eq:SRQ-recover-j-b} follows
from $\mathbf{t}_{l,\mu}\bmod\Lambda_{A,j+1}=0$, for $\mu>j$, since
$\mathbf{t}_{l,\mu}\in\Lambda_{A,\mu}\cap\mathcal{V}_{A,\mu+1}$;
\eqref{eq:SRQ-recover-j-c} follows from $\mathbf{t}_{l,j}=\mathbf{0}$
for $l\notin\mathcal{S}^{\left\langle j\right\rangle }$ by the definition
of $\mathbf{t}_{l,j}$ and $\mathcal{S}^{\left\langle j\right\rangle }$.
Also by the definition of $\mathbf{t}_{l,j}$, we have $\mathbf{t}_{l,j}\in\mathcal{C}_{e}^{\left\langle j\right\rangle }=\Lambda_{A,j}\cap\mathcal{V}_{A,j+1}$.
In Line \ref{state:U-hat-q} of Algorithm \ref{alg:SRQ}, $\psi^{-\left\langle j\right\rangle }$
maps $\mathbf{t}_{l,j}$ from $\mathcal{C}_{e}^{\left\langle j\right\rangle }$
to the finite field $\mathbb{F}_{\gamma}^{k}$. Denote $\mathbf{w}_{l,j}=\psi^{-\left\langle j\right\rangle }\left(\mathbf{t}_{l,j}\right)$,
for $l=1,\cdots,L$. From \eqref{eq:equivalent-lattice-encoding-quan}
and Lemma 6 in \cite{nazer2011compute}, we obtain the following isomorphism:
\begin{equation}
\sum_{l\in\mathcal{S}^{\left(i\right)}}q_{ml}\mathbf{w}_{l,j}=\psi^{-\left\langle j\right\rangle }\left(\left[\sum_{l\in\mathcal{S}^{\left\langle j\right\rangle }}a_{ml}\mathbf{t}_{l,j}\right]\bmod\Lambda_{A,j+1}\right).
\end{equation}
Then by \eqref{eq:SRQ-recover-j} we have 
\begin{eqnarray}
\sum_{l\in\mathcal{S}^{\left(i\right)}}q_{ml}\mathbf{w}_{l,j} & = & \psi^{-\left\langle j\right\rangle }\left(\hat{\mathbf{v}}_{m}^{\left\langle j\right\rangle }\bmod\Lambda_{A,j+1}\right).\label{eq:message_equation-quan}
\end{eqnarray}
Stacking the two sides of \eqref{eq:message_equation-quan} for $m\in\mathcal{T}^{\left\langle j\right\rangle }$
in a row-by-row manner gives a matrix equation over $\mathbb{F}_{\gamma}$:
\begin{equation}
\mathbf{Q}^{\left\langle j\right\rangle }\mathbf{W}^{\left\langle j\right\rangle }=\mathbf{U}^{\left\langle j\right\rangle }\label{eq:U-QW-quan}
\end{equation}
where the rows in $\mathbf{W}^{\left\langle j\right\rangle }$ are
specified by $\mathbf{w}_{l,j}^{T},l\in\mathcal{S}^{\left\langle j\right\rangle }$
and $\mathbf{U}^{\left\langle j\right\rangle }$ is the matrix obtained
in \eqref{eq:SRQ-v_m} of Algorithm \ref{alg:SRQ}. Recall that $\mathbf{Q}^{\left\langle j\right\rangle }$
is of full rank by assumption. We obtain from \eqref{eq:U-QW-quan}
that 
\begin{equation}
\mathbf{W}^{\left\langle j\right\rangle }=\left(\mathbf{Q}^{\left\langle j\right\rangle }\right)^{-1}\mathbf{U}^{\left\langle j\right\rangle }=\hat{\mathbf{W}}^{\left\langle j\right\rangle }.\label{eq:W-QU-quan}
\end{equation}
Thus, we can obtain $\mathbf{w}_{l,j},l\in\mathcal{S}^{\left\langle j\right\rangle }$
from \eqref{eq:W-QU-quan}. Then 
\begin{equation}
\psi^{\left\langle j\right\rangle }\left(\mathbf{w}_{l,j}\right)=\mathbf{t}_{l,j}=\hat{\mathbf{t}}_{l,j},l\in\mathcal{S}^{\left\langle j\right\rangle }.\label{eq:proof-hat-t=00003Dt}
\end{equation}
Note that $\mathbf{w}_{l,j}^{T}$ is in general not the $l$-th row
of $\mathbf{W}^{\left\langle j\right\rangle }$.

Then we cancel the contribution of $\left\{ \sum_{\mu=\pi_{c}\left(l\right)}^{j}\mathbf{t}_{l,\mu},l\in\mathcal{S}^{\left\langle j\right\rangle }\right\} $
from $\left\{ \mathbf{v}_{m},m\in\mathcal{T}^{\left\langle j+1\right\rangle }\right\} $,
yielding \eqref{eq:v_m_j+1} for $m\in\mathcal{T}^{\left\langle j+1\right\rangle }$,
where \eqref{eq:v_m_j+1_b} is from \eqref{eq:SRQ-v_m} and \eqref{eq:v_m-exp-only-quan};
\eqref{eq:v_m_j+1_c} and \eqref{eq:v_m_j+1_d} utilizes the fact
that, for $\mathbf{x}_{1}\in\mathbb{R}^{n},\mathbf{x}_{2}\in\Lambda$,
\begin{subequations}\label{eq:q(x1+x2)=00003Dq(x1)+x2} 
\begin{eqnarray}
Q_{\Lambda}\left(\mathbf{x}_{1}+\mathbf{x}_{2}\right) & = & \mathbf{x}_{1}+\mathbf{x}_{2}-\left(\mathbf{x}_{1}+\mathbf{x}_{2}\right)\bmod\Lambda\\
 & = & Q_{\Lambda}\left(\mathbf{x}_{1}\right)+\mathbf{x}_{2}
\end{eqnarray}
\end{subequations}and \eqref{eq:v_m_j+1_end} follows by noting 
\[
Q_{\Lambda_{d,m}}\negmedspace\left(\sum_{l=1}^{L}a_{ml}\sum_{\mu=1}^{j}\mathbf{t}_{l,\mu}\right)=Q_{\Lambda_{d,m}}\left(\sum_{l\in\mathcal{S}^{\left\langle j\right\rangle }}\negmedspace a_{ml}\negmedspace\sum_{\mu=\pi_{c}\left(l\right)}^{j}\mathbf{t}_{l,\mu}\right)
\]
by the definition of $\mathcal{S}^{\left\langle j\right\rangle }$
in \eqref{eq:S<j>} and $\mathbf{t}_{l,\mu}=0$ for $\mu<\pi_{c}\left(l\right)$.
Eq. \eqref{eq:v_m_j+1} shows that \eqref{eq:quantized-v_m^i-after-split}
holds for the $\left(j+1\right)$-th iteration as well. Thus, \eqref{eq:quantized-v_m^i-after-split}
holds by induction.

\begin{figure*}[t]
\begin{subequations}\label{eq:v_m_j+1} 
\begin{eqnarray}
 &   & \mathbf{v}_{m}^{\left\langle j+1\right\rangle }\\
 & = & \left[Q_{\Lambda_{d,m}}\negthickspace\left(\sum_{l=1}^{L}\negmedspace a_{ml}\negthickspace\left(\negmedspace\sum_{\mu=1}^{L}\mathbf{t}_{l,\mu}\negmedspace-\negmedspace Q_{\Lambda_{s}}\left(\mathbf{t}_{l}-\mathbf{d}_{l}\right)\right)\negthickspace\right)\negmedspace-\negmedspace Q_{\Lambda_{d,m}}\negthickspace\left(\sum_{l\in\mathcal{S}^{\left\langle j\right\rangle }}\negmedspace a_{ml}\negthickspace\sum_{\mu=\pi_{c}\left(l\right)}^{j}\negmedspace\mathbf{t}_{l,\mu}\right)\right]\negthickspace\bmod\Lambda_{s}\label{eq:v_m_j+1_b}\\
 & = & \left[\negmedspace Q_{\Lambda_{d,m}}\negthickspace\left(\sum_{l=1}^{L}\negmedspace a_{ml}\negthickspace\sum_{\mu=1}^{L}\negmedspace\mathbf{t}_{l,\mu}\negmedspace\right)\negmedspace-\negthickspace\sum_{l=1}^{L}\negmedspace a_{ml}Q_{\Lambda_{s}}\negmedspace\left(\mathbf{t}_{l}-\mathbf{d}_{l}\right)\negmedspace-\negmedspace Q_{\Lambda_{d,m}}\negthickspace\left(\negthickspace\sum_{l\in\mathcal{S}^{\left\langle j\right\rangle }}\negthickspace a_{ml}\negthickspace\sum_{\mu=\pi_{c}\left(l\right)}^{j}\negthickspace\mathbf{t}_{l,\mu}\right)\negmedspace\right]\negthickspace\bmod\Lambda_{s}\label{eq:v_m_j+1_c}\\
 & = & \left[Q_{\Lambda_{d,m}}\left(\sum_{l=1}^{L}a_{ml}\sum_{\mu=1}^{L}\mathbf{t}_{l,\mu}\right)-Q_{\Lambda_{d,m}}\left(\sum_{l\in\mathcal{S}^{\left\langle j\right\rangle }}a_{ml}\sum_{\mu=\pi_{c}\left(l\right)}^{j}\mathbf{t}_{l,\mu}\right)\right]\bmod\Lambda_{s}\\
 & = & \left[Q_{\Lambda_{d,m}}\negmedspace\left(\sum_{l=1}^{L}\negmedspace a_{ml}\negmedspace\sum_{\mu=1}^{j}\negmedspace\mathbf{t}_{l,\mu}\right)\negmedspace+\negmedspace\sum_{l=1}^{L}\negmedspace a_{ml}\negmedspace\sum_{\mu=j+1}^{L}\negmedspace\mathbf{t}_{l,\mu}\negmedspace-\negmedspace Q_{\Lambda_{d,m}}\left(\negmedspace\sum_{l\in\mathcal{S}^{\left\langle j\right\rangle }}\negmedspace a_{ml}\negthickspace\sum_{\mu=\pi_{c}\left(l\right)}^{j}\negmedspace\mathbf{t}_{l,\mu}\right)\right]\negmedspace\bmod\Lambda_{s}\label{eq:v_m_j+1_d}\\
 & = & \left[\sum_{l=1}^{L}a_{ml}\sum_{\mu=j+1}^{L}\mathbf{t}_{l,\mu}\right]\bmod\Lambda_{s}\label{eq:v_m_j+1_end}
\end{eqnarray}
\end{subequations}
\hrulefill
\vspace*{4pt}
\end{figure*}

Since Algorithm \ref{alg:SRQ} recovers $\left\{ \mathbf{t}_{l,j},l\in\mathcal{S}^{\left\langle j\right\rangle }\right\} $
by \eqref{eq:proof-hat-t=00003Dt} for $j=1,\cdots,L$, and $\mathbf{t}_{l,j}=\mathbf{0}$
for $l\notin\mathcal{S}^{\left\langle j\right\rangle }$, we see that
all $\left\{ \mathbf{t}_{l,j}\right\} $ are recovered. Finally, Algorithm
\ref{alg:SRQ} recovers $\left\{ \mathbf{t}_{l}\right\} $ by \eqref{eq:lattice-codeword-split-simplified},
which concludes the proof.

Furthermore, from \eqref{eq:proof-hat-t=00003Dt} and $\mathbf{t}_{l,j}=\mathbf{0}$
for $l\notin\mathcal{S}^{\left\langle j\right\rangle }$, we see that
\[
\hat{\mathbf{t}}_{l}=\left[\sum_{\mu=\pi_{c}\left(j\right)}^{L}\hat{\mathbf{t}}_{l,\mu}\right]\bmod\Lambda_{s}=\left[\sum_{\mu=\pi_{c}\left(l\right)}^{L}\mathbf{t}_{l,\mu}\right]\bmod\Lambda_{s}=\mathbf{t}_{l}
\]
which concludes the proof.

\section{Proof of Lemma \ref{lem:pi_d-design}\label{sec:proof-pi_d-design}}

By assumption, $\mathbf{Q}^{\left\langle L\right\rangle }=\mathbf{Q}$
is of full rank. From \eqref{eq:S<j>} and \eqref{eq:T<j>}, we see
that $\mathbf{Q}^{\left\langle j-1\right\rangle }$ can be obtained
by deleting one row (corresponding to the $\pi_{d}^{-1}\left(j\right)$-th
relay) and one column (corresponding to the $\pi_{c}^{-1}\left(j\right)$-th
source) from $\mathbf{Q}^{\left\langle j\right\rangle }$.

Suppose $\mathbf{Q}^{\left\langle j\right\rangle }\in\mathbb{F}_{\gamma}^{j\times j},j>1,$
is of full rank, and we want to find a $\mathbf{Q}^{\left\langle j-1\right\rangle }\in\mathbb{F}_{\gamma}^{\left(j-1\right)\times\left(j-1\right)}$
of full rank. Since $\pi_{c}\left(\cdot\right)$ is given, we delete
the corresponding column of $\mathbf{Q}^{\left\langle j\right\rangle }$,
yielding a matrix $\tilde{\mathbf{Q}}^{\left\langle j\right\rangle }$
of rank $\left(j-1\right)$. From linear algebra, there always exists
at least one full-rank $\mathbf{Q}^{\left\langle j-1\right\rangle }$
obtained by deleting one row of $\tilde{\mathbf{Q}}^{\left\langle j\right\rangle }$.
Choose such a $\mathbf{Q}^{\left\langle j-1\right\rangle }$ (with
rank $\left(j-1\right)$) and set the corresponding value of $\pi_{d}\left(\cdot\right)$
according to the index of the deleted row. By induction, we obtain
full-rank $\mathbf{Q}^{\left\langle L\right\rangle },\cdots,$ and
$\mathbf{Q}^{\left\langle 1\right\rangle }$.

\section{Proof of Lemma \ref{lem:mod-lattice-decodability}\label{sec:Proof-of-SRM}}

We follow Algorithm \ref{alg:SRM} step by step to prove that the
output of Algorithm \ref{alg:SRM} satisfies $\hat{\mathbf{t}}_{l}=\mathbf{t}_{l},l=1,\cdots,L$.
Algorithm \ref{alg:SRM} involves $L$ iterations. In the $i$-th
iteration, $\mathbf{t}_{\pi_{s}^{-1}\left(i\right)}$ is recovered
with $\left\{ \mathbf{v}_{m}^{\left(i\right)},m\in\mathcal{T}^{\left(i\right)}\right\} $,
where $\mathbf{v}_{m}^{\left(i\right)}$ is the lattice equation from
the $m$-th relay at the $i$-th iteration.

We first prove that, in the $i$-th iteration, 
\begin{equation}
\mathbf{v}_{m}^{\left(i\right)}=\left[\sum_{l\in\mathcal{S}^{\left(i\right)}}a_{ml}\left(\mathbf{t}_{l}-Q_{\Lambda_{s,l}}\left(\mathbf{t}_{l}-\mathbf{d}_{l}\right)\right)\right]\bmod\Lambda_{e,m}\label{eq:v_j^i-expression}
\end{equation}
for $m\in\mathcal{T}^{\left(i\right)}$, and that exactly one of $\left\{ \mathbf{t}_{l}\right\} $
is restored in each iteration. We prove this statement by induction.
The algorithm sets $\mathbf{v}_{m}^{\left(1\right)}=\mathbf{v}_{m},m\in\mathcal{T}^{\left(1\right)}$,
at the initialization stage. We immediately see that \eqref{eq:v_j^i-expression}
holds for $i=1$. Now suppose that \eqref{eq:v_j^i-expression} holds
for the $i$-th iteration. We will show that $\mathbf{t}_{\pi_{s}^{-1}\left(i\right)}$
is recovered in the $i$-th iteration, and that \eqref{eq:v_j^i-expression}
also holds for the $\left(i+1\right)$-th iteration.

By the definition of $\mathcal{S}^{\left(i\right)}$ and $\mathcal{T}^{\left(i\right)}$,
$\Lambda_{B,i}=\Lambda_{s,\pi_{s}^{-1}\left(i\right)}$ is the finest
in $\left\{ \Lambda_{s,l}\vert l\in\mathcal{S}^{\left(i\right)}\right\} $,
and also the finest in $\left\{ \Lambda_{e,m}\vert m\in\mathcal{T}^{\left(i\right)}\right\} $.
Then, with $\left\{ \mathbf{v}_{m}^{\left(i\right)}\right\} $, we
obtain 
\begin{eqnarray}
 &  & \mathbf{v}_{m}^{\left(i\right)}\bmod\Lambda_{B,i}\nonumber \\
 & = & \left[\sum_{l\in\mathcal{S}^{\left(i\right)}}a_{ml}\left(\mathbf{t}_{l}-Q_{\Lambda_{s,l}}\left(\mathbf{t}_{l}-\mathbf{d}_{l}\right)\right)\right]\bmod\Lambda_{B,i}\nonumber \\
 & = & \left[\sum_{l\in\mathcal{S}^{\left(i\right)}}a_{ml}\left(\mathbf{t}_{l}\bmod\Lambda_{B,i}\right)\right]\bmod\Lambda_{B,i}\label{eq:v_j-equivalent-lattice}
\end{eqnarray}
where $\mathbf{t}_{l}\bmod\Lambda_{B,i}$ is a codeword in the effective
lattice codebook $\mathcal{C}_{e}^{\left(i\right)}$ defined in Subsection
\ref{sub:SRM-alg}. That is, $\mathbf{t}_{l}\bmod\Lambda_{B,i}$ can
be seen as from a common codebook $\mathcal{C}_{e}^{\left(i\right)}$.
Note that 
\begin{equation}
\mathbf{t}_{\pi_{s}^{-1}\left(i\right)}\bmod\Lambda_{B,i}=\mathbf{t}_{\pi_{s}^{-1}\left(i\right)}\label{eq:t_pi_c_i_equation_relation}
\end{equation}
since $\Lambda_{B,i}=\Lambda_{s,\pi_{s}^{-1}\left(i\right)}$.

Define $\mathbf{w}_{l}^{\left(i\right)}=\psi^{-\left(i\right)}\left(\mathbf{t}_{l}\bmod\Lambda_{B,i}\right)$,
for $l=1,\cdots,L$. By \eqref{eq:v_j-equivalent-lattice}, \eqref{eq:equivalent-lattice-encoding},
and Lemma 6 in \cite{nazer2011compute}, we have 
\begin{equation}
\sum_{l\in\mathcal{S}^{\left(i\right)}}q_{ml}\mathbf{w}_{l}^{\left(i\right)}=\psi^{-\left(i\right)}\left(\mathbf{v}_{m}^{\left(i\right)}\bmod\Lambda_{B,i}\right).\label{eq:message_equation}
\end{equation}
Form the matrix $\mathbf{W}^{\left(i\right)}$ by stacking $\mathbf{w}_{l}^{\left(i\right)T},l\in\mathcal{S}^{\left(i\right)}$,
row by row. Similarly, form the matrix $\mathbf{U}^{\left(i\right)}$
by stacking $\left(\psi^{-\left(i\right)}\left(\mathbf{v}_{m}^{\left(i\right)}\bmod\Lambda_{B,i}\right)\right)^{T},m\in\mathcal{T}^{\left(i\right)}$.
We can write \eqref{eq:message_equation} as 
\begin{equation}
\mathbf{Q}^{\left(i\right)}\mathbf{W}^{\left(i\right)}=\mathbf{U}^{\left(i\right)},\label{eq:U-QW}
\end{equation}
which is a matrix equation over $\mathbb{F}_{\gamma}^{k}$. Then 
\begin{equation}
\mathbf{W}^{\left(i\right)}=\left(\mathbf{Q}^{\left(i\right)}\right)^{-1}\mathbf{U}^{\left(i\right)}=\hat{\mathbf{W}}^{\left(i\right)}\label{eq:W=Q_invU}
\end{equation}
where $\mathbf{Q}^{\left(i\right)}$ is of full rank by assumption.
From \eqref{eq:t_pi_c_i_equation_relation}, we have $\mathbf{w}_{\pi_{s}^{-1}\left(i\right)}=\mathbf{w}_{\pi_{s}^{-1}\left(i\right)}^{\left(i\right)}$.
By setting $\hat{\mathbf{w}}_{\pi_{s}^{-1}\left(i\right)}^{T}$ as
the corresponding row of $\hat{\mathbf{W}}^{\left(i\right)}$ in Line \ref{state:recover},
we recover the message as $\hat{\mathbf{w}}_{\pi_{s}^{-1}\left(i\right)}=\mathbf{w}_{\pi_{s}^{-1}\left(i\right)}$,
and the lattice codeword as 
\[
\hat{\mathbf{t}}_{\pi_{s}^{-1}\left(i\right)}=\mathbf{t}_{\pi_{s}^{-1}\left(i\right)}.
\]
That is, the lattice codeword of the $\pi_{s}^{-1}\left(i\right)$-th
source is correctly recovered.

Then the destination cancels the contribution of $\mathbf{t}_{\pi_{s}^{-1}\left(i\right)}$
and $\mathbf{d}_{\pi_{s}^{-1}\left(i\right)}$ from $\mathbf{v}_{m}^{\left(i\right)}$
in Line \ref{state:construct-v_i+1} of Algorithm \ref{alg:SRM} to
obtain $\mathbf{v}_{m}^{\left(i+1\right)}$. By noting $\mathcal{S}^{\left(i+1\right)}=\mathcal{S}^{\left(i\right)}\setminus\left\{ \pi_{s}^{-1}\left(i\right)\right\} $
and $\mathcal{T}^{\left(i+1\right)}=\mathcal{T}^{\left(i\right)}\setminus\left\{ \pi_{e}^{-1}\left(i\right)\right\} $,
we have 
\begin{equation}
\mathbf{v}_{m}^{\left(i+1\right)}\negmedspace=\negmedspace\left[\negmedspace\sum_{l\in\mathcal{S}^{\left(i+1\right)}}\negthickspace a_{ml}\left(\mathbf{t}_{l}-Q_{\Lambda_{s,l}}\left(\mathbf{t}_{l}-\mathbf{d}_{l}\right)\right)\right]\negthickspace\bmod\negmedspace\Lambda_{e,m}\label{eq:v_j-induction}
\end{equation}
for $m\in\mathcal{T}^{\left(i+1\right)}$, which establishes \eqref{eq:v_j^i-expression}
by induction. This concludes the proof of Theorem \ref{lem:mod-lattice-decodability}.

\section{Proof of Lemma \ref{lem:pi_e-design}\label{sec:proof-pi_e-design}}

By assumption, $\mathbf{Q}^{\left(1\right)}=\mathbf{Q}$ is of full
rank. From \eqref{eq:S(i)} and \eqref{eq:T(i)}, we see that $\mathbf{Q}^{\left(i+1\right)}$
can be obtained by deleting one row (corresponding to the $\pi_{e}^{-1}\left(j\right)$-th
relay) and one column (corresponding to the $\pi_{s}^{-1}\left(j\right)$-th
source) from $\mathbf{Q}^{\left(i\right)}$.

Suppose $\mathbf{Q}^{\left(i\right)}\in\mathbb{F}_{\gamma}^{\left(L-i+1\right)\times\left(L-i+1\right)},i>1,$
is of full rank, and we want to find a $\mathbf{Q}^{\left(i+1\right)}\in\mathbb{F}_{\gamma}^{\left(L-i\right)\times\left(L-i\right)}$
of full rank. Since $\pi_{s}\left(\cdot\right)$ is given, we delete
the corresponding column of $\mathbf{Q}^{\left(i\right)}$, yielding
a matrix $\tilde{\mathbf{Q}}^{\left(i\right)}$ of rank $\left(L-i\right)$.
From linear algebra, there always exists at least one full-rank $\mathbf{Q}^{\left(i+1\right)}$
obtained by deleting one row of $\tilde{\mathbf{Q}}^{\left(i\right)}$.
Choose such a $\mathbf{Q}^{\left(i+1\right)}$ (with rank $\left(L-i\right)$)
and set the corresponding value of $\pi_{e}\left(\cdot\right)$ according
to the index of the deleted row. By induction, we can obtain $\mathbf{Q}^{\left(1\right)},\cdots,\mathbf{Q}^{\left(L\right)}$
that are all of full rank.

\section{Proof of Lemma \ref{lem:quan-mod-lattice-decodability}\label{sec:Proof-of-SRMQ}}

In Section \ref{sec:Asymmetric-Quan-Mod}, we have shown that the
SRQ algorithm can be applied on $\left\{ \mathbf{v}_{m}\bmod\Lambda_{B,1}\right\} $
to recover $\left\{ \mathbf{t}_{l}\bmod\Lambda_{B,1}\right\} $. We
next show that the SRM algorithm can be used to recover $\left\{ Q_{\Lambda_{B,1}}\left(\mathbf{t}_{l}\right)\right\} $.
From Line \ref{state:joint-SRQ} of the SRMQ algorithm, we have \begin{subequations}\label{eq:t-quan-exp}
\begin{eqnarray}
\hat{\mathbf{t}}_{l}^{\textrm{quan}} & = & \left[\sum_{\mu=1}^{L}\mathbf{t}_{l,\mu}\right]\bmod\Lambda_{B,1}\\
 & = & \left[\sum_{\mu=1}^{L}\mathbf{t}_{l,\mu}+\mathbf{t}_{l,L+1}\right]\bmod\Lambda_{B,1}\label{eq:t-quan-exp-b}\\
 & = & \mathbf{t}_{l}\bmod\Lambda_{B,1}\label{eq:t-quan-exp-c}
\end{eqnarray}
\end{subequations}where step \eqref{eq:t-quan-exp-b} follows from
$\mathbf{t}_{l,L+1}\in\Lambda_{B,1}$, step \eqref{eq:t-quan-exp-c}
from \eqref{eq:lattice-codeword-split-general}. In Line \ref{state:v_m-quan},
we cancel $\left\{ \hat{\mathbf{t}}_{l}^{\textrm{quan}}\right\} $
from $\left\{ \mathbf{v}_{m}\right\} $, yielding \begin{subequations}\label{eq:v_m-SRMQ}
\begin{align}
 & \mathbf{v}_{m}^{\textrm{quan}}\\
= & \left[\mathbf{v}_{m}-Q_{\Lambda_{d,m}}\left(\sum_{l=1}^{L}a_{ml}\hat{\mathbf{t}}_{l}^{\textrm{quan}}\right)\right]\bmod\Lambda_{e,m}\\
= & \left[\sum_{l=1}^{L}a_{ml}\left(\mathbf{t}_{l}-\hat{\mathbf{t}}_{l}^{\textrm{quan}}-Q_{\Lambda_{s,l}}\left(\mathbf{t}_{l}-\mathbf{d}_{l}\right)\right)\right]\bmod\Lambda_{e,m}\label{eq:v_m-quan-b}\\
= & \left[\negmedspace\sum_{l=1}^{L}\negmedspace a_{ml}\negmedspace\left(Q_{\negthinspace\Lambda_{\negthinspace B\negthinspace,1}}\negthickspace\left(\mathbf{t}_{l}\right)\negmedspace-\negmedspace Q_{\negthinspace\Lambda_{\negthinspace s\negthinspace,l}}\negthickspace\left(\negthinspace Q_{\negthinspace\Lambda_{\negthinspace B\negthinspace,1}}\negthickspace\left(\mathbf{t}_{l}\right)\negmedspace-\negmedspace\left(\negthinspace\mathbf{d}_{l}\negmedspace-\negmedspace\hat{\mathbf{t}}_{l}^{\textrm{quan}}\negthinspace\right)\right)\right)\negmedspace\right]\negthickspace\bmod\negthickspace\Lambda_{\negthinspace e\negthinspace,m}\label{eq:v_m-quan-end}
\end{align}
\end{subequations}where \eqref{eq:v_m-quan-b} follows from \eqref{eq:v_m-exp-quan-mod}
and \eqref{eq:q(x1+x2)=00003Dq(x1)+x2}, \eqref{eq:v_m-quan-end}
from the fact of $\hat{\mathbf{t}}_{l}^{\textrm{quan}}=\mathbf{t}_{l}\bmod\Lambda_{B,1}$.
Note that $\mathbf{v}_{m}^{\textrm{quan}}$ in \eqref{eq:v_m-quan-end}
is actually a special case of \eqref{eq:v_m-exp-only-mod} with $\mathbf{t}_{l}$
replaced by $Q_{\Lambda_{B,1}}\left(\mathbf{t}_{l}\right)$, and $\mathbf{d}_{l}$
replaced by $\mathbf{d}_{l}^{\textrm{quan}}=\mathbf{d}_{l}-\hat{\mathbf{t}}_{l}^{\textrm{quan}}$,
for $l=1,\cdots,L$. Therefore, we can apply the SRM algorithm to
$\left\{ \mathbf{v}_{m}^{\textrm{quan}}\right\} $, yielding outputs
$\left\{ \hat{\mathbf{t}}_{l}^{\textrm{mod}}=Q_{\Lambda_{B,1}}\left(\mathbf{t}_{l}\right)\right\} $.
Finally, we obtain \begin{subequations} 
\[
\hat{\mathbf{t}}_{l}=\hat{\mathbf{t}}_{l}^{\textrm{quan}}+\hat{\mathbf{t}}_{l}^{\textrm{mod}}=\mathbf{t}_{l}\bmod\Lambda_{B,1}+Q_{\Lambda_{B,1}}\left(\mathbf{t}_{l}\right)=\mathbf{t}_{l}
\]
for $l=1,\cdots,L$, \end{subequations}which follows from \eqref{eq:mod-quantization-relation}
and \eqref{eq:t-quan-exp}.

\ifCLASSOPTIONcaptionsoff
  \newpage
\fi

\bibliographystyle{IEEEtran}
\bibliography{CF}



%





\end{document}